\documentclass[12pt]{article}

\usepackage{authblk}
\usepackage{tabularx}
\usepackage[margin=1in]{geometry}
\usepackage{amsmath,amssymb,graphicx,color}
\usepackage[titletoc,title]{appendix}
\usepackage{bm,mathrsfs}

\newcommand{\hzero}{h^{(0)}}
\newcommand{\hone}{h^{(1)}}
\newcommand{\htwo}{h^{(2)}}
\newcommand{\hthree}{h^{(3)}}

\newcommand{\xone}{x^{(1)}}
\newcommand{\xtwo}{x^{(2)}}
\newcommand{\xthree}{x^{(3)}}

\newcommand{\Hkern}{\mathcal{H}}

\newcommand{\sdone}[2]{\frac{\mathrm{d} #1}{\mathrm{d} #2}}

\newcommand{\pd}[2]{\frac{\partial #1}{\partial #2}}
\newcommand{\pdd}[3]{\frac{\partial^{#1} #2}{\partial #3^{#1}}}

\newcommand{\Me}{M\kern-.15em e \,}
\newcommand{\cc}{\mathrm{c}.\mathrm{c}.}
\newcommand{\mye}{\text{e}}
\newcommand{\myi}{\text{i}}

\usepackage{amsthm}

\newcommand{\matt}[1]{\textcolor{red}{#1}}

\title{A discrete complex Ginzburg-Landau equation for a hydrodynamic active lattice}
\author[1,2]{Stuart J. Thomson}
\author[2]{Matthew Durey}
\author[2]{Rodolfo R. Rosales}
\affil[1]{School of Engineering, Brown University, Providence, RI 02912, USA}
\affil[2]{Department of Mathematics, Massachusetts Institute of Technology, Cambridge, MA 02139, USA}
\date{}
\begin{document}
\maketitle

\begin{abstract}
A discrete and periodic complex Ginzburg-Landau equation, coupled to a discrete mean equation, is systematically derived from a driven and dissipative oscillator model, close to the onset of a supercritical Hopf bifurcation. The oscillator model is inspired by recent experiments exploring active vibrations of quasi-one-dimensional lattices of self-propelled millimetric droplets bouncing on a vertically vibrating fluid bath. Our systematic derivation provides a direct link between the constitutive properties of the lattice system and the coefficients of the resultant amplitude equations, paving the way to compare the emergent nonlinear dynamics---namely discrete bright and dark solitons, breathers, and traveling waves---against experiments. Further, the amplitude equations allow us to rationalize the successive bifurcations leading to these distinct dynamical states. The framework presented herein is expected to be applicable to a wider class of oscillators characterized by the presence of a dynamic coupling potential between particles. More broadly, our results point to deeper connections between nonlinear oscillators and the physics of active and driven matter.
\end{abstract}


\section{Introduction} 
The celebrated complex Ginzburg-Landau equation (CGLE) \cite{aranson2002world,garcia2012complex} is a generic model describing the dynamics of spatially extended, dissipative systems near a Hopf bifurcation. 
In contrast to the potentially complex, high-dimensional microscopic equations regulating a particular physical system, amplitude equations \cite{cross1993pattern,newell1993order} such as the CGLE are typically cast in terms of only a few macroscopic variables, or order parameters \cite{coullet1989defect, tanaka2003complex, garcia2008nonlocal, denk2016active, tan2020topological}. In general, the form of such effective models may be posited on phenomenological grounds, their structure determined through a combination of linear stability and symmetry arguments \cite{cross1993pattern}. This universal approach can, however, obfuscate the connection between the coefficients of the amplitude equation and the physical parameters of the system under study. A more robust approach sacrifices derivational simplicity in favor of obtaining the amplitude equation(s) \emph{directly} from the underlying microscopic equations of the system, typically continuous nonlinear partial differential equations \cite{newell1969finite, segel1969distant, cross1980derivation, stoop2015curvature}. However, for systems which are fundamentally discrete---for example, nonlinear oscillators---amplitude equations are typically posited as discretized versions of their continuous counterparts, seldom derived in a systematic manner from the original governing equations \cite{hakim1992dynamics,ravoux2000stability,maruno2003exact}.

We herein present a rigorous framework to systematically derive a discrete and periodic complex Ginzburg-Landau equation (dCGLE) for a driven and dissipative nonlinear oscillator, close to the onset of a supercritical Hopf bifurcation. The oscillator model is inspired by recent experiments exploring the active vibrations of a hydrodynamic lattice of self-propelled millimetric droplets \cite{thomson2020collective,thomson2020hydro}. The coefficients appearing in our dCGLE are directly related to the constitutive properties of the physical lattice system, paving the way to compare the numerical results of the resultant amplitude equations---namely the emergence of discrete bright and dark solitons, breathers, and travelling waves---against experiments. Further, the amplitude equations allow us to rationalize the successive bifurcations leading to these distinct dynamical states. Although we present the case of the hydrodynamic lattice, we propose that the framework presented herein is applicable to a wider class of oscillators characterized by the presence of a \emph{dynamic} coupling potential between particles. On a fundamental level, our results suggest deeper connections between nonlinear and nonlocal oscillators and the physics of active and driven matter \cite{denk2016active,sethia2014chimera,thakur2019collective}.  

\subsection{The hydrodynamic active lattice}
\begin{figure}
\begin{center}
\includegraphics[width=\textwidth]{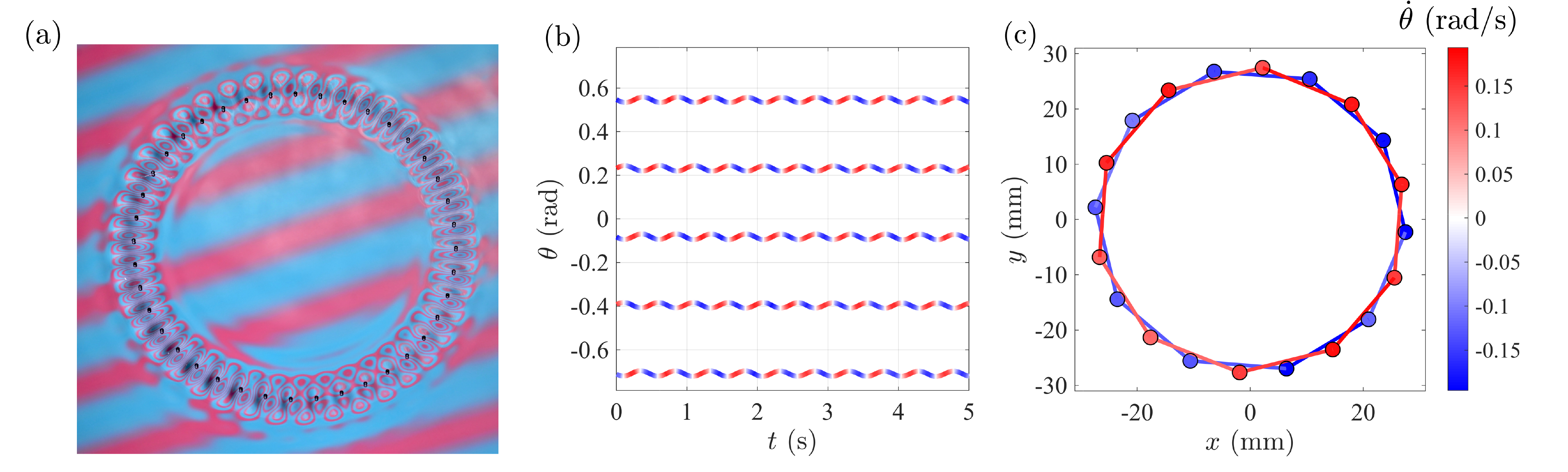}
\end{center}
\caption{(a) Overhead perspective of a chain of 40 equispaced, millimetric droplets of silicone oil, confined to an annular channel and surrounded by a shallow layer of fluid. The reflected color in the channel emphasises the deformation of the fluid surface as droplets impact the bath and excite subcritical Faraday waves. (b) A subset of droplet polar positions obtained from experiments for a lattice consisting of 20 droplets \cite{thomson2020collective}. Each droplet undergoes out-of-phase oscillations with respect to its neighbor, following a supercritical (Hopf) bifurcation \cite{thomson2020hydro}. (c) The instantaneous positions of all 20 droplets in the lattice for the same experiment as (b). The net result of the instability is the out-of-phase oscillations of two decahedral sub-lattices, colored red and blue.}
\label{fig:fig1}
\end{figure}

This study is motivated by experiments of quasi-one-dimensional lattices of millimetric droplets, bouncing synchronously and periodically on the surface of a vertically vibrating fluid bath and confined to an annular channel \cite{thomson2020collective}; see Figure \ref{fig:fig1}. (For a broader perspective of the physics of bouncing droplets, see \cite{couder2005walking, bush2015pilot, bush2018introduction} and references therein.) Upon successive impacts, each droplet excites a field of standing waves whose decay time, $T_{M}$, increases with the vertical acceleration of the bath and diverges at the Faraday threshold \cite{eddi2011information,molavcek2013walking}. The superposition of the wave fields generated by each droplet forms the global lattice wave field, which acts as an inter-droplet potential, mediating the spatiotemporal coupling of the lattice. This wave-mediated coupling represents a distinguishing feature of this new class of coupled oscillator: the waves produced at each droplet impact give rise to an effective self-generated, \emph{dynamic} coupling potential between droplets, one that evolves continuously with the droplet motion. 

For sufficiently weak vibrational forcing, the droplets exhibit stationary bouncing in a circular, equispaced lattice. Above a critical vertical acceleration of the bath (or, alternatively, critical decay time, $T_{M}$), the droplets destabilize to small lateral perturbations, oscillating about their equilibrium position (Figures 1b and 1c). Physically, these oscillations emerge due to the competition between droplet self-propulsion---arising through the propulsive force enacted on each droplet by the local slope of the lattice wave field---and wave-mediated, nonlocal coupling between droplets. Oscillations of the lattice are further offset by dissipative effects due to drag. That self-propulsion is achieved and sustained by the continual exchange of energy of the droplet with its environment---in this case, the vibrating bath---renders this hydrodynamic lattice a novel example of an inertial active system \cite{bechinger2016active, scholz2018inertial, lim2019cluster, lowen2020inertial}.

As shown in experiments \cite{thomson2020collective}, oscillations of the lattice follow the onset of either a supercritical or subcritical Hopf bifurcation, depending primarily on the proximity of neighboring droplets. Our focus here is on the supercritical case, for which periodic, small-amplitude, out-of-phase oscillations arise beyond the bifurcation point, initially uniform over all droplets. (When the bifurcation is subcritical, the dynamics are profoundly different: in experiments, the system approaches a distant attractor and the emergent dynamics manifest as a self-sustaining, nonlinear solitary-like wave \cite{thomson2020collective}.) The dependence of the form of these bifurcations on the parameters of the lattice system, and the ensuing dynamics of the uniform, periodic state, was characterized \emph{via} a weakly nonlinear analysis of a mathematical model describing the droplet lattice \cite{thomson2020hydro}. Upon further increase of the vibrational forcing, this periodic state can itself destabilize, leading to spatial modulations of the droplet oscillation amplitude, a phenomenon not captured by the analysis presented in \cite{thomson2020hydro}. To explore and rationalize the onset and resultant dynamics of these spatial modulations, we present a generalized weakly nonlinear theory of the lattice, in the vicinity of the supercritical Hopf bifurcation. We proceed to briefly summarize the results of \cite{thomson2020hydro} as they pertain to our derivation of the governing amplitude equations presented herein.

\subsection{Lattice model and linear theory}
\label{sec:model_and_linear}
\begin{figure}[h!]
\begin{center}
\includegraphics[width=\textwidth]{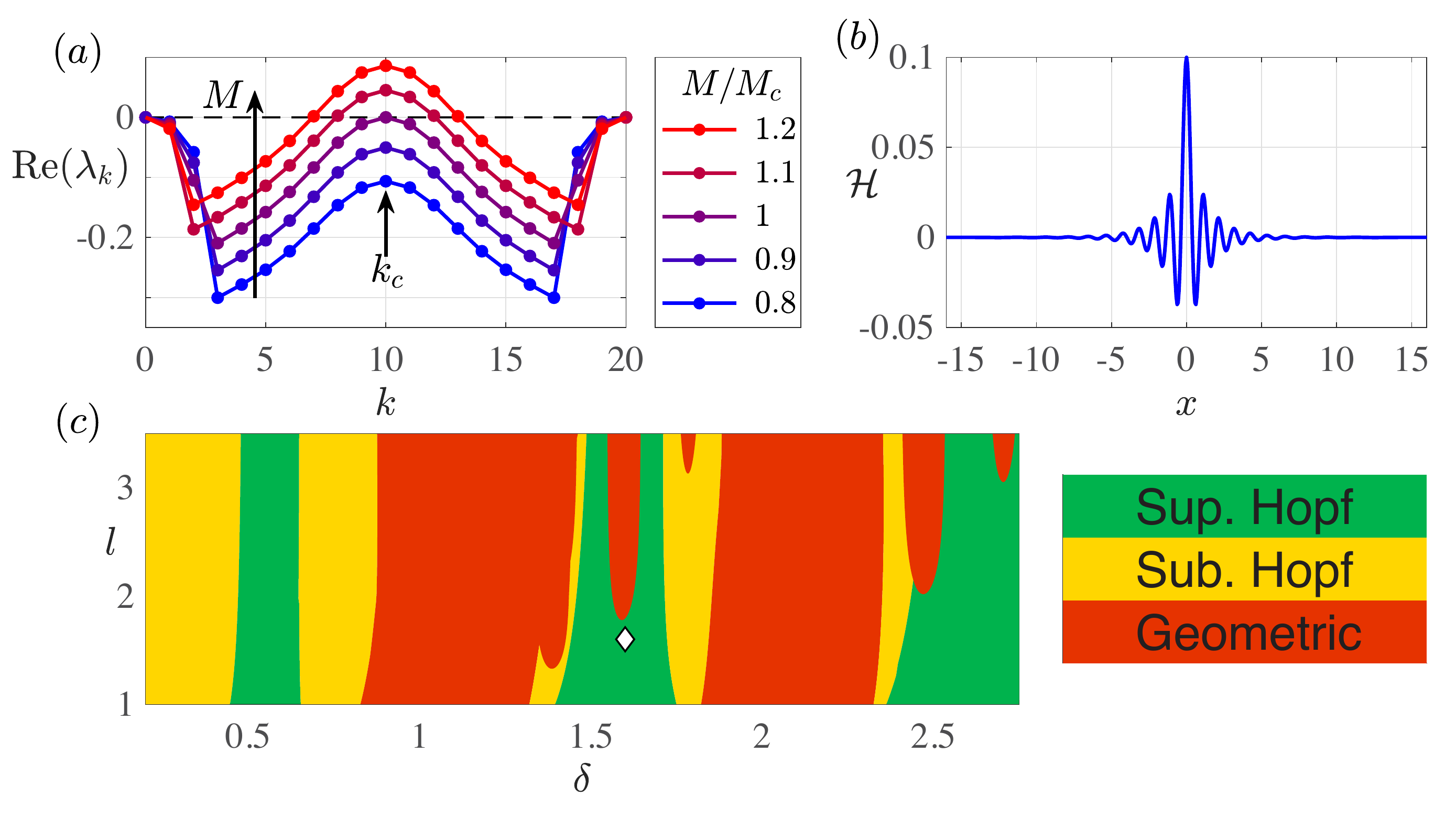}
\end{center}
\caption{
(a) Behavior of the eigenvalues of \eqref{eqn:disp_relation} in the case of a supercritical Hopf bifurcation for $N = 20$ droplets. At a critical threshold $M = M_{c}$, a single wavenumber $k_{c} = 10$ first touches $\text{Re}(\lambda_{k_{c}}) = 0$.
(b) The prototypical wave-field kernel, $\mathcal{H}(x)$, used in numerical simulations presented herein (see \S\ref{sec:change_wave} for details). 
(c) Regime diagram summarising the stability of an equispaced lattice of $N = 20$ droplets as the parameters $l$ and $\delta$ are varied independently \cite{thomson2020collective,thomson2020hydro}, as derived from \eqref{eqn:M_disprelation} and \eqref{eqn:Hopf_TDR}. We delineate regimes of super- and sub-critical Hopf bifurcations, as well as geometric frustration of the equispaced lattice.  The diamond indicates the parameter values used in (a) and (b), specifically $\delta = l = 1.6$. 
}
\label{fig:RD}
\end{figure}

\emph{Model}---The principal assumption underpinning the hydrodynamic lattice model \cite{thomson2020hydro} is that the horizontal motion of each of the $N$ droplets in the lattice may be averaged over one bouncing period, which we denote $T_{F}$. This \emph{stroboscopic approximation} \cite{oza2013trajectory} effectively eliminates the droplets' synchronous vertical motion from consideration. To further simplify matters, we assume that the droplets lie on a circle of constant radius, $R$, which, in experiments, is determined by the inner and outer radii of the annular channel. Combining this motion with the stroboscopic approximation yields the following equation of motion for the circumferential position, $x_{n}(t)$, of each droplet in the lattice \cite{thomson2020hydro}:

\begin{subequations}
\label{eqn:drops_EoM_dim}
\begin{equation}
\label{eqn:drops_EoM_dim1}
m \ddot{x}_{n} + \bar{D}\dot{x}_{n} = -m g \frac{\partial h}{\partial x}\bigg\vert_{x = x_{n}}.
\end{equation}
Dots denote differentiation with respect to time, $t$, and the space variable, $x\in[0,L = 2\pi R]$, is directed along the circumference of the circle on which the droplets lie. According to equation \eqref{eqn:drops_EoM_dim1}, the motion of each droplet of mass $m$ is thus governed by a balance between inertia, a linear drag with drag coefficient $\bar{D}$, and the time-averaged propulsive wave force enacted on each droplet by the local slope of the lattice wave field, $h(x,t)$. By periodicity, $h(x,t) = h(x + L,t)$. The remaining parameter, $g$, is acceleration due to gravity. It is to be understood that $h(x,t)$ is the stroboscopic \emph{global} lattice wave field---the time-averaged \emph{superposition} of wave fields generated by each individual droplet in the lattice---projected onto the circle. 

A distinguishing feature of the hydrodynamic lattice is that the propulsive wave force enacted on each droplet depends explicitly on the prior trajectory of \emph{every} droplet in the lattice \cite{thomson2020hydro,eddi2011information, molavcek2013walking, oza2013trajectory}. The time-dependent evolution of the lattice wave field, $h$, may be described by
\begin{equation}
\label{eqn:drops_EoM_dim2}
\frac{\partial h}{\partial t} + \frac{1}{T_{M}}h = \frac{1}{T_{F}}\sum_{m = 1}^{N}\Hkern(x - x_{m}),
\end{equation}
where the wave kernel, $\Hkern(x)$, is the quasistatic wave field generated by stationary bouncing of each individual droplet, time-averaged over $T_{F}$ and projected onto the circle \cite{thomson2020hydro}.  Prompted by the fundamental aspects of the fluid system \cite{thomson2020collective}, our theory requires only that $\mathcal{H}(x)$ be sufficiently smooth and periodic with $\Hkern(x) = \Hkern(x + L)$, exhibiting variations over a characteristic wavelength, $\lambda$, and an exponential spatial decay with lengthscale $l_d$; see Figure \ref{fig:RD}(b) for a prototypical example. (A candidate wave kernel satisfying these properties is presented in \S\ref{sec:BF_dyn}.) In summary, equation \eqref{eqn:drops_EoM_dim2} represents a balance between the rate-of-change of $h$, wave dissipation, and the superposition of waves generated about the instantaneous position of each droplet in the lattice.
\end{subequations}

The dynamical system \eqref{eqn:drops_EoM_dim} is non-dimensionalized \emph{via} the scalings 
$$
t = \frac{m}{\bar{D}}\hat{t} = t_{0}\hat{t},\qquad x = \lambda \hat{x},\qquad h = \frac{\lambda^2}{g t_{0}^2}\hat{h} = h_{0}\hat{h},\qquad \Hkern = \frac{h_{0}T_{F}}{t_{0}}\hat{\Hkern}.
$$
Upon dropping the carets, we arrive at the dimensionless system \cite{thomson2020hydro}
\begin{subequations}
\label{eqn:drops_nondim}
\begin{align}
\label{eqn:drops_nondim_1}
\ddot{x}_{n} + \dot{x}_{n} &= -\frac{\partial h}{\partial x}(x_{n},t),\\
\label{eqn:drops_nondim_2}
\frac{\partial h}{\partial t} + \nu h &= \sum_{m = 1}^{N}\Hkern(x - x_{m}),
\end{align}
\end{subequations}
where $\nu = t_0/T_M$ is the dimensionless dissipation rate of the wave field. Recalling that the decay time, $T_{M}$, may be regarded as a proxy for the vertical vibrational acceleration of the bath \cite{eddi2011information,molavcek2013walking}, $\nu$ plays the role of a control, or bifurcation, parameter in the dimensionless system \eqref{eqn:drops_nondim}. While $\nu$ is convenient for algebraic manipulations, we will interpret our results in terms of the dimensionless \emph{memory parameter}, $M = 1/\nu$, where the influence of prior droplet dynamics increases with $M$. For future reference, we provide a list of the salient dimensionless parameters related to the lattice model \eqref{eqn:drops_nondim} in Table \ref{table:param_table}.

\begin{table}
\begin{center}
\begin{tabularx}{\textwidth}{ p{0.325\textwidth} | p{0.625\textwidth} }
\hline
Parameter & Definition\\
\hline
Lattice model \eqref{eqn:drops_nondim} & \\
$x_n$ & droplet positions\\
$N$ & number of droplets\\
$h$ & stroboscopic lattice wave field\\
$\Hkern$ & single-droplet quasistatic wave kernel with characteristic wavelength, $\lambda$, decay length, $l_d$, and amplitude, $\mathcal{A}$\\
$\alpha = 2\pi/N$; $\delta = \alpha R /\lambda = \alpha r_0$ & droplet angular separation; droplet arc-length separation \\
$\nu$; $M = 1/\nu$ & dissipation rate of wave field; memory parameter \\
$\varepsilon = \sqrt{\nu_c - \nu}$; $\nu_c = 1/M_c$ & perturbation parameter; instability threshold for supercritical Hopf bifurcation\\
$k_c$; $\omega_c$ & critical wavenumber and angular frequency at supercritical Hopf bifurcation\\
\hline
Amplitude equations \eqref{eqn:dCGLE_intro} & \\
$\mu = \alpha/\varepsilon$; $\mu_c$ & bifurcation parameter for amplitude equations; threshold for onset of spatial modulations\\
$A_n$, $B_n$ & $n$th droplet oscillation amplitude and rotational drift\\
$c_g$; $\sigma_i$; $\gamma_i$; $D_i$ & group speed; growth coefficients; coupling coefficients; diffusion coefficients \\
\hline
\end{tabularx}
\end{center}
\caption{List of salient parameters in the lattice model \eqref{eqn:drops_nondim}, amplitude equations \eqref{eqn:dCGLE_intro}, and stability analysis thereof.}
\label{table:param_table}
\end{table}

\emph{Linear theory}---We consider a static, equispaced lattice with droplet positions $x_n = n\delta$ and $h = \nu^{-1}\sum_{m = 1}^{N} \Hkern(x - m\delta)$, where $\delta = 2\pi R/\lambda N$ is the droplet arc-length separation along the circle of dimensionless radius $R/\lambda$. The critical value of $M$ above which the wave force promotes sustained self-propulsion of the droplets, and hence oscillations of the lattice, is determined from the linear stability of the lattice system  \eqref{eqn:drops_nondim}. We summarize the key results; full details are presented in \cite{thomson2020hydro}. 

Coinciding with experiments \cite{thomson2020collective}, we consider small perturbations of the droplet positions of the form
\begin{equation}
\label{eqn:xn_pert}
x_{n}(t) = n\delta + \eta\left[A\exp(\myi k n \alpha + \lambda_{k}t) + \cc\right],\quad \eta\ll 1,
\end{equation}
with a concomitant perturbation to the wave field, $h$.
Here $\alpha = 2\pi/N$ is the angular spacing of the droplets, $A$ is a complex amplitude, $\myi$ is the imaginary unit, and $\cc$ denotes complex conjugation of the preceding term. The eigenvalues, $\lambda_{k}$, for each integer wavenumber, $k$, satisfy the dispersion relation $\mathcal{D}_{k}(\lambda_{k}; \nu) = 0$ \cite{thomson2020hydro}, where
\begin{equation}
\label{eqn:disp_relation}
\mathcal{D}_{k}(\lambda_{k}; \nu) = \lambda^2_{k} + \lambda_{k} + \frac{c_{0}}{\nu} - \frac{c_{k}}{\lambda_{k} + \nu}
\end{equation}
and the real constants $c_{k}$ are defined as
$$
c_{k} = \sum_{n = 1}^{N}\cos(kn\alpha)\Hkern''(n\delta).
$$
By symmetry considerations, we need only consider discrete wavenumbers in the interval $k\in [0,\overline{N}]$, where $\overline{N} = \lfloor N/2\rfloor$. Notably, the coefficients $c_{k}$ depend on both the number of droplets, $N$, and the droplet separation, $\delta$. 

After rearranging $\mathcal{D}_{k}(\lambda_{k}; \nu) = 0$ and writing $\nu = 1/M$, the eigenvalues, $\lambda_{k}$, describing the asymptotic linear stability of the equispaced lattice are roots of the cubic polynomial $q_k(\lambda_k; M)$, where
\begin{equation}
\label{eqn:M_disprelation}
q_k(\lambda_k; M) = M \lambda^{3}_{k} + (M + 1)\lambda^{2}_{k} + (c_0 M^2 + 1)\lambda_k + M(c_0 - c_k).
\end{equation}
As the memory parameter, $M$, is increased, the lattice becomes asymptotically unstable if, for some wavenumber, $k$, there is at least one eigenvalue, $\lambda_{k}$, for which $\text{Re}(\lambda_{k}) > 0$. We note, however, that a fundamental property of the lattice system is its rotational invariance, characterized by the neutrally stable eigenvalue $\lambda_{0} = 0$. As we shall see in \S\ref{sec:idea_summary}, this invariance gives rise to a discrete mean equation coupled to our dCGLE describing the rotational drift of the lattice.

As discussed in \cite{thomson2020hydro}, the stability of the lattice system, whose eigenvalues are the roots of \eqref{eqn:M_disprelation}, depends in a non-trivial way on the droplet separation parameter, $\delta$, through the coefficients $c_{k}$. Specifically, the lattice can destabilize \emph{via} two distinct mechanisms, depending primarily on the droplet separation, $\delta$: (i) an oscillatory instability, wherein the real part of a pair of complex-conjugate eigenvalues transitions from negative to positive as $M$ increases through $M = M_c = 1/\nu_c$, or (ii) a so-called ``geometrical instability,'' independent of the memory parameter, $M$, brought on by geometrical frustration of the lattice wave field. We focus our attention in this paper on case (i), which arises when $c_k \le c_0$ for all $k$ \cite{thomson2020hydro}. In Figure \ref{fig:RD}(a), we present an oscillatory instability arising for $N = 20$ droplets: as the control parameter, $M$, is increased, a single wavenumber $k_{c} = N/2$ first touches $\text{Re}(\lambda_{k_{c}}) = 0$ for some critical value of $M = M_{c}$, while $\text{Im}(\lambda_{k_{c}}) = \omega_{c}\neq 0$, corresponding to an oscillatory instability of the lattice. In terms of the coefficients $c_k$ and critical memory, $M_c$, the angular frequency, $\omega_c$, satisfies $\omega^2_c = c_0 M_c + 1/M_c$ \cite{thomson2020hydro}. Further, that the system destabilizes to a pair of complex-conjugate eigenvalues points to a Hopf bifurcation \cite{strogatz2018nonlinear}. 

The existence of a Hopf bifurcation was confirmed in \cite{thomson2020hydro} by an accompanying weakly nonlinear stability analysis of the equispaced lattice. The analysis presented in \cite{thomson2020hydro} demonstrates that, just beyond the instability threshold ($0 < M - M_c\ll 1$), each droplet evolves according to  
\begin{equation}
\label{eqn:sl_expansion}
x_n = n\delta + \left[D(T) + O(\varepsilon)\right] + \varepsilon\left[A(T)\text{e}^{\myi(k_c n \alpha + \omega_c t)} + \cc\right] + O(\varepsilon^2),
\end{equation}
where $0 < \varepsilon = \sqrt{\nu_c - \nu}\ll 1$. The slowly varying complex amplitude $A(T)$, a function of the slow timescale $T = \varepsilon^2 t$, is governed by a Stuart-Landau equation
\begin{subequations}
\label{eqn:Hopf_TDR}
\begin{equation}
\label{eqn:Hopf_StuLand}
\sdone{A}{T} = \sigma_1 A - \bar{\sigma}_2 |A|^2 A
\end{equation}
with an accompanying equation governing the evolution of the rotational drift, $D$, of the lattice, namely
\begin{equation}
\sdone{D}{T} = \bar{\gamma}_3 |A|^2.
\end{equation}
\end{subequations}
The origin of this rotational drift may be traced back to the $k = 0$ mode of the dispersion relation \eqref{eqn:M_disprelation}, corresponding to rotational invariance. As alluded to earlier, we will find an analogous equation in the amplitude equations presented in \S\ref{sec:idea_summary}. 

The complex coefficients $\sigma_1$, $\bar{\sigma}_2$, and $\bar{\gamma}_3$ in equations \eqref{eqn:Hopf_TDR} are defined in terms of the parameters of the lattice system \eqref{eqn:drops_nondim} \cite{thomson2020hydro}. We shall see how they are related to the coefficients of the amplitude equations derived in this paper in \S\ref{sec:idea_summary}. For now, we note that, consistent with the linear instability of the lattice system \eqref{eqn:drops_nondim}, the coefficient $\sigma_{1}$ satisfies $\text{Re}(\sigma_{1}) > 0$, corresponding to exponential growth of the oscillation amplitude. Whether the Hopf bifurcation is super- or sub-critical depends on the sign of $\text{Re}(\bar{\sigma}_2)$: For $\text{Re}(\bar{\sigma}_2) > 0$, the bifurcation is supercritical, and subcritical when $\text{Re}(\bar{\sigma}_2) < 0$. Further, when $k_c = N/2$, it is found that $\bar{\gamma}_3 = 0$, in which case $D = \text{constant}$, corresponding to an arbitrary shift of the droplet equilibrium positions.

The stability of the equispaced lattice and its complex dependence on the system parameters, is summarized in Figure \ref{fig:RD}(c). Each region of parameter space is color-coded according to the type of instability that arises for a lattice of $N = 20$ droplets as $\delta$ and the dimensionless spatial decay length of the wave kernel, $l = l_{d}/\lambda$, are varied independently. Of particular interest here are the green regions, indicating a supercritical Hopf bifurcation. A crucial feature of the stability analysis just described is the value of the critical wavenumber, $k_{c}$ \cite{thomson2020hydro}. When the Hopf bifurcation is supercritical, it is found that $k_{c} = \lfloor N/2 \rfloor$ (except near the boundaries with subcritical Hopf bifurcations), a fact that we will exploit in \S\ref{sec:idea_summary}. Further, when $N$ is odd, a symmetry-breaking, oscillatory-rotary motion of the lattice arises, which will manifest in \S\ref{sec:idea_summary} as a non-zero group speed in our dCGLE.

Numerical simulations of the hydrodynamic active lattice \eqref{eqn:drops_nondim} beyond the supercritical Hopf bifurcation (when $\nu < \nu_c$ or, equivalently, $M > M_c$) reveal the onset of a second bifurcation where spatially uniform, small-amplitude oscillations of the droplet positions give way to spatio-temporal modulations in the droplet oscillation amplitude, arising \emph{via} a long-wavelength instability. To capture this second bifurcation, and the ensuing dynamics, the weakly nonlinear analysis leading to the amplitude equations \eqref{eqn:Hopf_TDR} must be generalized to account for spatial, as well as temporal, effects. 

\subsection{The amplitude equations governing the lattice vibrations}
\label{sec:idea_summary}
Starting from the lattice system \eqref{eqn:drops_nondim}, we use the asymptotic method of multiple scales \cite{kevorkian2012multiple} to derive a generalized set of amplitude equations, accounting for both spatial and temporal modulations of the droplet oscillation amplitude and rotational drift. Specifically, we show that each droplet position evolves according to 
\begin{equation}
\label{eqn:dcgle_expansion}
x_{n} = n\delta + \varepsilon\left[A_{n}(T)\text{e}^{\myi (k_{c}n\alpha + \omega_{c}t)}+ \cc + B_{n}(T)\right] + O(\varepsilon^2),
\end{equation}
where $A_n$ is the slowly varying complex amplitude of the $n$th droplet in the lattice, $B_n$ is the rotational drift, and $T = \varepsilon^2 t$ is the slow time scale. As is typical in the method of multiple scales,  eliminating secular terms at successive orders of $\varepsilon$ yields a coupled system of equations for $A_{n}$ and $B_{n}$, resulting in a dCGLE coupled to a discrete mean equation:
\begin{subequations}
\label{eqn:dCGLE_intro}
\begin{equation}
\label{eqn:dCGLE_intro1}
\sdone{A_n}{T} + \mu^2 c_g\nabla A_{n} = \sigma_{1} A_n - \sigma_2 |A_n|^2 A_n + \mu\gamma_{1}A_{n}\nabla B_{n} + \mu^2 D_1\Delta A_{n},
\end{equation}
\begin{equation}
\label{eqn:dCGLE_intro2}
\sdone{B_n}{T} = \mu^2 D_2\Delta B_{n}
 + 2\mu\text{Re}\left[\gamma_{2}A_{n}\nabla A^{*}_{n}\right] + \mu\gamma_{3}|A_{n}|^2.
\end{equation}
\end{subequations}
The operators $\nabla$ and $\Delta$, defined in \S\ref{sec:WNL}, are discrete gradient and Laplacian operators, respectively. In terms of the dissipation rate, $\nu$, and the number of droplets, $N$, the control parameter, $\mu$, in the amplitude equations \eqref{eqn:dCGLE_intro} that arises from our analysis is defined
$$\mu = \frac{\alpha}{\varepsilon} = \frac{2\pi }{N \sqrt{\nu_c - \nu } }. $$
Our theory is valid for $\mu = O(1)$ or, equivalently, $\alpha\sim\varepsilon$. (We note that this condition requires that $N$ is sufficiently large.) Spatially uniform oscillations for which $A_{n}(T) = A(T)$ and $B_n(T) = B(T)$ are also captured by our theory, in which case the amplitude equations \eqref{eqn:dCGLE_intro} reduce in form to the Stuart-Landau equations \eqref{eqn:Hopf_TDR} with $D = \varepsilon B$. Table \ref{table:param_table} provides a reference list of relevant parameters appearing in the amplitude equations \eqref{eqn:dCGLE_intro}.

We pause to emphasize a few aspects of the system \eqref{eqn:dCGLE_intro}. With regards to the coefficients, the group speed, $c_g$, growth coefficients, $\sigma_{i}$ $(i = 1,2)$, coupling coefficients, $\gamma_i$ ($i = 1,2,3$), and the diffusion coefficients, $D_i$ ($i = 1,2$), are determined as part of the multiple scales analysis and are directly related to the physical parameters of the lattice system \eqref{eqn:drops_nondim} (see Appendix \ref{sec:sl_derive} for their algebraic forms). Notably, $D_2 > 0$ and $\gamma_3$ are both real, while the remaining parameters are all complex with $\text{Re}(\sigma_1) > 0$ and $\text{Re}(D_1) > 0$. Notably, $\sigma_2 = \bar{\sigma}_2 + O(\alpha)$, which is consistent with \eqref{eqn:Hopf_StuLand} as $\alpha\rightarrow 0$ and the absence of spatial effects. Further, a keen eye will note that the drift term, $B_n$, appears at $O(\varepsilon)$ in the expansion \eqref{eqn:dcgle_expansion}, whereas the drift is an $O(1)$ term in \eqref{eqn:sl_expansion}. As discussed in Appendix \ref{sec:sl_derive}, this change arises since $\bar{\gamma}_3 = \alpha\gamma_3$ when $k_c \lesssim \lfloor N/2 \rfloor$ and $\alpha\sim\varepsilon \ll 1$, resulting in $\gamma_3 = O(1)$; hence, the drift, $D$, appearing in \eqref{eqn:sl_expansion} is promoted to higher order, specifically $O(\varepsilon)$.

We note that, in our system, discreteness originates at the level of the underlying microscopic equations \eqref{eqn:drops_nondim}, and thus is a connate characteristic of the resultant amplitude equations \eqref{eqn:dCGLE_intro}. This feature is in contrast with discrete versions of the CGLE motivated by a direct discretization of the continuous CGLE using standard finite difference operators \cite{hakim1992dynamics,ravoux2000stability, maruno2003exact}. Similarly, periodicity arises from the periodicity of the lattice, rather than being imposed \emph{ex post facto} through periodic boundary conditions \cite{doering1988low,bartuccelli1990possibility}. Interestingly, the system \eqref{eqn:dCGLE_intro} is the discrete and periodic analogue of the amplitude equations describing a host of disparate physical phenomena presented elsewhere \cite{coullet1985propagative, matthews2000pattern, komarova2000nonlinear, cox2001new}.

\subsection{Outline}
This paper is organized as follows. For the interested reader, we provide further details of the multiple scales procedure involved in deriving the amplitude equations \eqref{eqn:dCGLE_intro} from the lattice model \eqref{eqn:drops_nondim} in \S\ref{sec:WNL}. We then proceed to analyze the successive bifurcations of the system \eqref{eqn:dCGLE_intro} in \S\ref{sec:BF_dyn}, rationalizing the onset of the second bifurcation leading to a long-wavelength instability of time-periodic oscillations of the lattice, and eventually to spatiotemporal chaos. Numerical solutions of the amplitude equations \eqref{eqn:dCGLE_intro} beyond the second bifurcation are presented in \S\ref{sec:numerical_solns}, demonstrating their rich dynamical behavior in the form of traveling waves, bright and dark solitons, and dark breathers, prompting future comparison with experiments. A review of our results is presented in \S\ref{sec:concl}, along with a discussion of future directions.


\section{Mechanics of the weakly nonlinear analysis}
\label{sec:WNL}
The weakly nonlinear analysis leading to the derivation of the amplitude equations \eqref{eqn:dCGLE_intro} assumes slowly varying spatial and temporal modulations of the oscillation amplitude of each droplet. For a supercritical Hopf bifurcation, stable, small-amplitude oscillations arise when $\nu$ is only slightly below the critical threshold, $\nu_{c}$ (corresponding to $M$ slightly above $M_{c}$); specifically we consider $0 < \varepsilon = \sqrt{\nu_c - \nu} \ll 1$ and small perturbations from the equispaced lattice configuration $x_{n} = n\delta$ and $h = \nu^{-1}_{c}\sum_{m = 1}^{N}\Hkern(x - m\delta)$. We then pose the following asymptotic multiple-scales expansions
\begin{equation}
\label{eqn:ms_expansions}
x_{n} \sim n\delta + \sum_{i=1}^{\infty}\varepsilon^{i}x^{(i)}_{n}(t,T),\quad h \sim \frac{1}{\nu_{c}}\sum_{m=1}^{N}\Hkern(x - m\delta) + \sum_{i=1}^{\infty} \varepsilon^{i}h^{(i)}(x,t,T),
\end{equation}
where the slow time-scale is $T = \varepsilon^2 t$. 

Following the typical recipe for the method of multiple scales, inserting \eqref{eqn:ms_expansions} into the lattice system \eqref{eqn:drops_nondim} leads to a hierarchy of problems for $x^{(i)}_{n}$ and $h^{(i)}$ at successive orders of $\varepsilon$. A series of solvability conditions must then be satisfied at each order of $\varepsilon$ to guarantee the suppression of secular terms that would otherwise lead to unbounded solutions and a violation of the multiple-scales \emph{ansatz}. Satisfying the solvability condition arising at $O(\varepsilon^3)$ leads to the amplitude equations \eqref{eqn:dCGLE_intro}. Before arriving there, however, there are several aspects of the current problem that complicate the weakly nonlinear, multiple-scales analysis of the lattice system \eqref{eqn:drops_nondim}. Our derivation falls into three stages: (i) we first solve for the wave field perturbations, $h^{(i)}$, allowing us to project the wave force onto the droplet trajectories, $x^{(i)}_n$; (ii) we next exploit the spatial decay of the wave kernel, $\Hkern$, and the assumed slowly varying spatial effects to approximate the nonlocal inter-droplet dynamics by $p$-nearest-neighbor interactions, where $p$ is determined by the decay length of the wave and the spacing of the lattice; (iii) finally, we consider the limit of weak asymmetry when the number of droplets is large (equivalently, when $k_c$ departs slightly from $N/2$). We elaborate on these three \emph{key ideas} below, with a full account provided in Appendix \ref{sec:sl_derive}.

\subsection{Solving for the wave field}
\label{sec:solve_wf}
Our first point of interest is at $O(\varepsilon)$, where we have the system
\begin{subequations}
\label{eqn:WNL_Oe}
\begin{align}
\label{eqn:WNL_Oe1}
\pdd{2}{\xone_{n}}{t} + \pd{\xone_{n}}{t} &= - \frac{\xone_{n}}{\nu_{c}}\sum_{m = 1}^{N}\Hkern''((n-m)\delta) -\pd{\hone}{x}\bigg\vert_{x = n\delta},\\
\label{eqn:WNL_Oe2}
\pd{\hone}{t} + \nu_{c}\hone &= -\sum_{m = 1}^{N}\xone_{m}\Hkern'(x - m\delta).
\end{align}
\end{subequations}
At this stage in conventional multiple-scales analyses of nonlinear oscillators \cite{strogatz2018nonlinear, kevorkian2012multiple}, one is typically interested in solving for the perturbed oscillator position, $\xone_{n}$, alone. In the present case, however, we must also contend with equation \eqref{eqn:WNL_Oe2} governing the free surface perturbation, $\hone$. In order to project the dynamics entirely onto the droplet trajectories, $\xone_{n}$, our \emph{first key idea} is to use the form of \eqref{eqn:WNL_Oe2} to define the auxiliary variables, $X_{n}$, satisfying \cite{thomson2020hydro}
\begin{equation}
\label{eqn:Xn_aux}
\pd{X_{n}}{t} + \nu_{c}X_{n} = \xone_{n}.
\end{equation}
A particular solution of \eqref{eqn:WNL_Oe2} is then
\begin{equation}
\label{eqn:WNL_hone}
\hone = -\sum_{m = 1}^{N}X_{m}\Hkern'(x - m\delta).
\end{equation}
We neglect the homogeneous solution of \eqref{eqn:WNL_Oe2}, which decays exponentially on the fast time-scale, $t$. Now that $\hone$ is expressed in terms of the auxiliary variables, $X_{n}$, through equation \eqref{eqn:WNL_hone}, the linear system \eqref{eqn:WNL_Oe} may be recast as a dynamical system for $\xone_{n}$ and $X_{n}$. Specifically, substituting \eqref{eqn:WNL_hone} into \eqref{eqn:WNL_Oe1} yields $\mathcal{L}_{n}\bm{x}^{(1)} = 0$, where $\bm{x}^{(1)} = \left(\xone_1,\ldots,x^{(1)}_{N}\right)$ and the linear operator, $\mathcal{L}_{n}$, is defined as
\begin{equation}
\label{eqn:xone_Lop}
\mathcal{L}_{n}\bm{x}^{(1)} = \pdd{2}{\xone_{n}}{t} + \pd{\xone_{n}}{t} + \sum_{m = 1}^{N}\left(\frac{\xone_{n}}{\nu_{c}} - X_{m}\right)\Hkern''(\delta(n-m)).
\end{equation}

Informed by the \emph{ansatz} \eqref{eqn:xn_pert}, we now seek a solution to $\mathcal{L}_{n}\bm{x}^{(1)} = 0$ of the form
\begin{equation}
\label{eqn:xone_ansatz}
\xone_{n} = A_{n}(T)\mye^{\myi (k_{c}n\alpha + \omega_{c}t)} + \cc + B_{n}(T),\quad X_{n} = \frac{A_{n}(T)}{\nu_{c} + \myi\omega_{c}}\mye^{\myi (k_{c}n\alpha + \omega_{c}t)} + \cc + \frac{1}{\nu_{c}}B_{n}(T),
\end{equation}
where we recall that the critical wavenumber of instability, $k_{c}$, and angular frequency, $\omega_{c}$, are determined from linear theory (\S\ref{sec:model_and_linear}). We note that the presence of the subscript $n$ in both the complex amplitude, $A_{n}$, and mean, $B_{n}$, generalizes the spatially uniform expansion \eqref{eqn:sl_expansion}, which simply leads to a Stuart-Landau equation \cite{thomson2020hydro}. By inserting \eqref{eqn:xone_ansatz} into \eqref{eqn:xone_Lop}, we find that
\begin{multline}
\label{eqn:AnmBnm}
\mathcal{L}_n \bm{x}^{(1)} = \left\{\frac{\mye^{\myi\phi_n}}{\nu_c + \myi \omega_c}\sum_{m = 1}^{N}(A_n - A_{n-m})\mye^{-\myi k_c m \alpha}\Hkern''(m\delta)\right\} + \cc \\ 
+ \frac{1}{\nu_c}\sum_{m=1}^{N} (B_n - B_{n-m})H''(m\delta),
\end{multline}
where $\phi_n = k_c n \alpha + \omega_c t$. We note that, \emph{en route} to obtaining \eqref{eqn:AnmBnm}, we first write $A_{n-m} = (A_{n-m} - A_n) + A_n$ and then simplify the resultant expression using the properties of the dispersion relation, namely $\mathcal{D}_k(\lambda_k; \nu_c) = \mathcal{D}_0(0; \nu_c) = 0$.

\subsection{Approximating discrete convolutions}
\label{sec:approx_convs}
We now arrive at the \emph{second key idea}, which lies at the heart of our analysis: approximating the discrete convolutions arising in equation \eqref{eqn:AnmBnm}. When $A_{n}$ and $B_{n}$ are spatially uniform (i.e.\ independent of $n$), the right-hand side of \eqref{eqn:AnmBnm} is identically zero. To allow for spatial variations, we approximately satisfy \eqref{eqn:AnmBnm} at this order by approximating the discrete convolutions. Recalling the angular spacing parameter, $\alpha = 2\pi/N$, we assume a distinguished limit between the relative sizes of $\alpha$ and $\varepsilon$. Specifically, we assume $\alpha \sim \varepsilon$ and thus set $\alpha = \mu\varepsilon$, where $\mu = O(1)$ is the control parameter arising in the amplitude equations \eqref{eqn:dCGLE_intro}. This limiting case allows us to approximate convolutions by spatially local terms.

\textcolor{black}{The following approximation technique applies to any $2\pi$-periodic function $F(\theta)$ that is slowly varying (that is, the derivatives of $F$ are all of size $O(1)$) and to any periodic function $G(x) = G(x + L)$ exhibiting exponential spatial decay. By denoting $F_m = F(m\alpha)$, we approximate convolutions of the form
\begin{equation}
\label{eq:conv_example}
\sum_{m=1}^N F_{n-m} G(m\delta),
\end{equation}
by first introducing an interpolating polynomial of degree $2p$ passing through the points $F_{n-p}, \ldots, F_{n+p}$.
As justified in Appendix \ref{sec:conv_approx}, for $|m| \leq p = O(1)$, we may then express 
\begin{equation}
\label{eqn:F_expansion}
F_{n-m} = F_n - \alpha m \nabla F_n + \alpha^2\frac{m^2}{2}\Delta F_n  + O(\alpha^3),
\end{equation}
where $\nabla$ and $\Delta$ are the central finite difference operators approximating the first and second derivatives of $F(\theta)$ using $2p + 1$ points spaced $\alpha$ apart. The difference stencils for $p = 1$ and $p = 2$ are listed in Appendix \ref{sec:conv_approx}.}

\textcolor{black}{
We then exploit the assumed exponential decay of $G$ to extend the polynomial approximation \eqref{eqn:F_expansion} outside of the interval $m = -p,\ldots, p$, only incurring exponentially small errors when substituting into the convolution, provided that $p$ is sufficiently large. It follows that the convolution \eqref{eq:conv_example} may be approximated by
\begin{multline}
\label{eqn:conv_expansion}
\sum_{m=1}^N F_{n-m} G(m\delta) =\\ F_n\sum_{m=1}^N G(m\delta) - \alpha \nabla F_n  \sum_{m=1}^N a_m G(m\delta) + \alpha^2 \Delta F_n  \sum_{m=1}^N b_m G(m\delta) + O(\alpha^3),
\end{multline}
where the periodic coefficients $a_m = a_{m + N}$ and $b_m = b_{m+N}$ are odd and even, respectively, defined as
$$
a_m = m\qquad\text{and}\quad b_m = \frac{1}{2}m^2\quad\text{for}\quad |m| < N/2,
$$
with $a_{N/2} = 0$ and $b_{N/2} = N^2/8$ for $N$ even.
}

\textcolor{black}{
The coupling integer $p > 0$ is chosen so that $|G(\pm p\delta) | \ll 1$; due to the exponential decay of $G$ over the length scale $l$, this condition is equivalent to $\exp(-p\delta/l) \ll 1$.
Moreover, if $p$ becomes too large then the neglected terms in \eqref{eqn:F_expansion} may also be large since $\alpha p$ may not be sufficiently small: hence, $p$ must also satisfy $\alpha p \ll 1$. For the case where $N$ is large ($\alpha \ll 1$) and $l \sim \delta$, $p = 1$ and 2 adequately satisfy both of these conditions. 
We note that the larger the value of $p$, the greater the level of coupling between droplets in the resultant amplitude equations \eqref{eqn:dCGLE_intro}, where $p = 1$ corresponds to nearest-neighbor interactions.}

Employing the slowly varying approximation \eqref{eqn:conv_expansion} in \eqref{eqn:AnmBnm} yields the sought-after approximation to the discrete convolutions. Specifically, we obtain
\begin{multline}
\label{eqn:disp_relation_expand}
\mathcal{L}_n\bm{x}^{(1)} + \left\{-\alpha\frac{\nabla A_{n}\mye^{\myi\phi_{n}}}{\nu_{c} + \myi\omega_{c}}\sum_{m = 1}^{N}a_{m}\mye^{-\myi k_{c}m\alpha}\Hkern''(m\delta) + \alpha^2\frac{\Delta A_{n}\mye^{\myi\phi_{n}}}{\nu_{c} + \myi\omega_{c}}\sum_{m=1}^{N}b_{m}\mye^{-\myi k_{c}m\alpha}\Hkern''(m\delta)\right\} + \cc\\
+ \alpha^2\frac{\Delta B_{n}}{\nu_{c}}\sum_{m = 1}^{N}b_{m}\Hkern''(m\delta) = O(\alpha^3).
\end{multline}
We note that there is not a $\nabla B_n$ term in \eqref{eqn:disp_relation_expand} as the symmetry of the wave kernel, $\Hkern$, determines that its coefficient vanishes, specifically $\sum_{m = 1}^{N}a_m \Hkern''(m\delta) = 0$. Recalling our assumption that $\alpha\sim \varepsilon$, terms of $O(\alpha^{n})$ in equations \eqref{eqn:disp_relation_expand} are consequently promoted to $O(\varepsilon^{n + 1})$, appearing as secular terms (either those constant in $t$ or proportional to $\mye^{\myi\phi_{n}}$) on the right-hand of the expansion of equation \eqref{eqn:drops_nondim_1}. Thus, the $\Delta A_n$ and $\Delta B_n$ terms in equation \eqref{eqn:disp_relation_expand} are destined to become the diffusion-like terms in the amplitude equations \eqref{eqn:dCGLE_intro}.

\subsection{The limit of weak asymmetry}
\label{sec:weak_asymm}
Finally, our \emph{third key idea} is to use the observed property of the supercritical Hopf bifurcations, outlined in Figure \ref{fig:RD}(a), that $k_{c} \lesssim \lfloor N/2 \rfloor$. We first define $\chi = N/2 - k_c$, where we typically find that $\chi = 0$ or $\chi = 1/2$. Then, by recalling that $\alpha = 2\pi/N$, we use the form $k_c = N/2 - \chi$ to recast the $\nabla A_n$ coefficient in \eqref{eqn:disp_relation_expand} as
$$
\frac{\alpha}{\nu_{c} + \myi\omega_{c}}\sum_{m = 1}^{N}a_{m}\mye^{-\myi k_{c}m\alpha}\Hkern''(m\delta) = \frac{\myi\alpha}{\nu_{c} + \myi\omega_{c}}\sum_{m=1}^{N} a_{m}(-1)^{m}\sin(m\alpha\chi)\Hkern''(m\delta),
$$
which is an even function of $\alpha$ for $\chi \neq 0$, and vanishes otherwise. In the former case, this term is expected to be of size $O(\alpha^2)$ as $\alpha\rightarrow 0$, which may be verified numerically. Further, we are prompted to define the $O(1)$ group velocity
$$
\hat{c}_g = \frac{1}{\alpha}\sum_{m = 1}^{N}\frac{a_{m}\mye^{-\myi k_{c}m\alpha}\Hkern''(m\delta)}{\nu_{c} + \myi\omega_{c}},
$$
which vanishes when $k_c = N/2$. We then acknowledge---as a corollary of our  second key idea---that the term $\alpha^2 \hat{c}_g\nabla A_n$ appears as a secular term at $O(\varepsilon^3)$, ultimately resulting in the advective term in our dCGLE \eqref{eqn:dCGLE_intro1}. (We note that $\hat{c}_g$ is divided by a further coefficient to yield the $c_g$ in \eqref{eqn:dCGLE_intro}; see Appendix \ref{sec:sl_derive} for clarification.) In this final step, equation \eqref{eqn:disp_relation_expand} reduces to $\mathcal{L}_n \bm{x}^{(1)}  = O(\varepsilon^3)$, verifying our \emph{ansatz} for $\xone_{n}$.

As shown in Appendix \ref{sec:sl_derive}, a combination of the foregoing three key ideas is used to systematically derive the amplitude equations \eqref{eqn:dCGLE_intro} from the lattice system \eqref{eqn:drops_nondim}. The remaining terms involving time derivatives and the nonlinear coupling terms in \eqref{eqn:dCGLE_intro} are obtained by eliminating higher-order secular terms, both those that are promoted from $O(\varepsilon^2)$ and others that appear at $O(\varepsilon^3)$. We now proceed to analyze the stability of the amplitude equations \eqref{eqn:dCGLE_intro}, elucidating the second bifurcation leading to spatiotemporal modulation of the droplet oscillation amplitude and drift as the control parameter, $\mu$, is decreased from infinity (corresponding to $\nu < \nu_c$).

\section{Stability of periodic oscillations}
\label{sec:BF_dyn}
As discussed in \S\S\ref{sec:model_and_linear} and \ref{sec:idea_summary}, we are concerned with the stability of the hydrodynamic lattice beyond the threshold of the supercritical Hopf bifurcation, specifically for $\nu < \nu_{c}$, or equivalently, for $\mu < \infty$. In this section, we elucidate the mechanism leading to a modulational instability of the spatially uniform solution of the amplitude equations \eqref{eqn:dCGLE_intro}. The onset of spatial amplitude modulations in the canonical complex Ginzburg-Landau equation is the eponymous Benjamin-Feir-Newell (BFN) instability, after its discovery in describing the instability of periodic surface gravity (Stokes) waves  \cite{benjamin1967disintegration, stuart1978eckhaus}. As we shall see, in our system, this instability takes a slightly different form due to the coupling of the complex amplitude, $A_{n}$, with the mean, $B_{n}$. In what follows, we conduct a linear stability of the amplitude equations \eqref{eqn:dCGLE_intro}, the computations for which are standard \cite{garcia2012complex}, but lengthy. We therefore highlight only the key features here. 
\subsection{Linear stability}
In the case of a supercritical Hopf bifurcation, where $\text{Re}(\bar{\sigma}_2) \sim \text{Re}(\sigma_{2}) > 0$ when $\alpha\ll 1$, there exists a spatially uniform solution to \eqref{eqn:dCGLE_intro} of the form
\begin{equation}
\label{eqn:dCGLE_periodic_sols}
A^{(0)}_{n}(T) = \rho\exp(\myi \Omega T)\quad\text{and}\quad B^{(0)}_{n}(T) = \mu\gamma_{3}\rho^2 T + \text{constant},
\end{equation}
where the modulus and angular frequency of the complex amplitude are
$$\rho = \sqrt{\text{Re}(\sigma_{1})/\text{Re}(\sigma_{2})}\quad\text{and}\quad\Omega = \text{Im}(\sigma_{1}\sigma^{*}_{2})/\text{Re}(\sigma_{2}).$$

We now consider small perturbations about the spatially uniform state of the form
$$
A_{n}(T) = A^{(0)}_{n}(T)(1 + \overline{A}_{n}(T)),\quad B_{n}(T) = B^{(0)}_{n}(T) + \overline{B}_{n}(T),
$$
where $0 < |\overline{A}_{n}|\sim |\overline{B}_{n}|\ll 1$. Substituting this \emph{ansatz} into \eqref{eqn:dCGLE_intro} and neglecting terms of quadratic order and higher, we obtain a linear system governing the perturbations $\overline{A}_{n}$ and $\overline{B}_{n}$, supplemented by an additional equation for $\overline{C}_{n} = \overline{A}^{*}_{n}$. The resulting linear system may then be diagonalized by considering a discrete Fourier transform in $n$. Specifically, we consider solutions of the form 
$$
\overline{A}_{n} = \sum_{\xi =0}^{N - 1}\hat{A}_{\xi}\mye^{\myi \xi n \alpha},\quad \overline{B}_{n} = \sum_{\xi =0}^{N - 1}\hat{B}_{\xi}\mye^{\myi \xi n \alpha},\quad \overline{C}_{n} = \sum_{\xi =0}^{N - 1}\hat{C}_{\xi}\mye^{\myi \xi n \alpha},
$$
where the \textcolor{black}{wavenumber, $\xi$, is an} integer. Under this transformation, we obtain 
\begin{equation}
\label{eqn:eigenvalue_system}
\sdone{\hat{\textbf{A}}_{\xi}}{T} = \text{M}_{\xi}(\mu)\hat{\textbf{A}}_{\xi},
\end{equation}
where $\hat{\textbf{A}}_{\xi} = (\hat{A}_{\xi},\hat{B}_{\xi},\hat{C}_{\xi})^{T}$ and
\begin{equation}
\label{eqn:fourier_matrix}
\text{M}_{\xi}(\mu) = \begin{pmatrix}
-\mu^2 c_g\nabla_{\xi} - \rho^2 \sigma_{2} + \mu^2 D_1\Delta_{\xi} & \mu\gamma_{1}\nabla_{\xi} & -\rho^2 \sigma_{2} \\
\mu\rho^2(\gamma^{*}_{2}\nabla_{\xi} + \gamma_{3}) & \mu^2 D_2\Delta_{\xi} & \mu\rho^2(\gamma_{2}\nabla_{\xi} + \gamma_{3}) \\
-\rho^2 \sigma^{*}_{2}  & \mu\gamma^{*}_{1}\nabla_{\xi} & -\mu^2c^{*}_{g}\nabla_{\xi} - \rho^2 \sigma^{*}_{2} + \mu^2 D^{*}_1\Delta_{\xi}
\end{pmatrix}.
\end{equation}
For the case of nearest-neighbor interactions ($p = 1$), the Fourier symbols, $\nabla_{\xi}$ and $\Delta_{\xi}$, of the difference operators, $\nabla$ and $\Delta$, are defined as
$$
\nabla_{\xi} = \frac{\myi\sin(\xi\alpha)}{\alpha}\quad\text{and}\quad \Delta_{\xi} = \frac{2(\cos(\xi\alpha) - 1)}{\alpha^2}.
$$

\begin{figure}[h!]
\begin{center}
\includegraphics[width=0.95\textwidth]{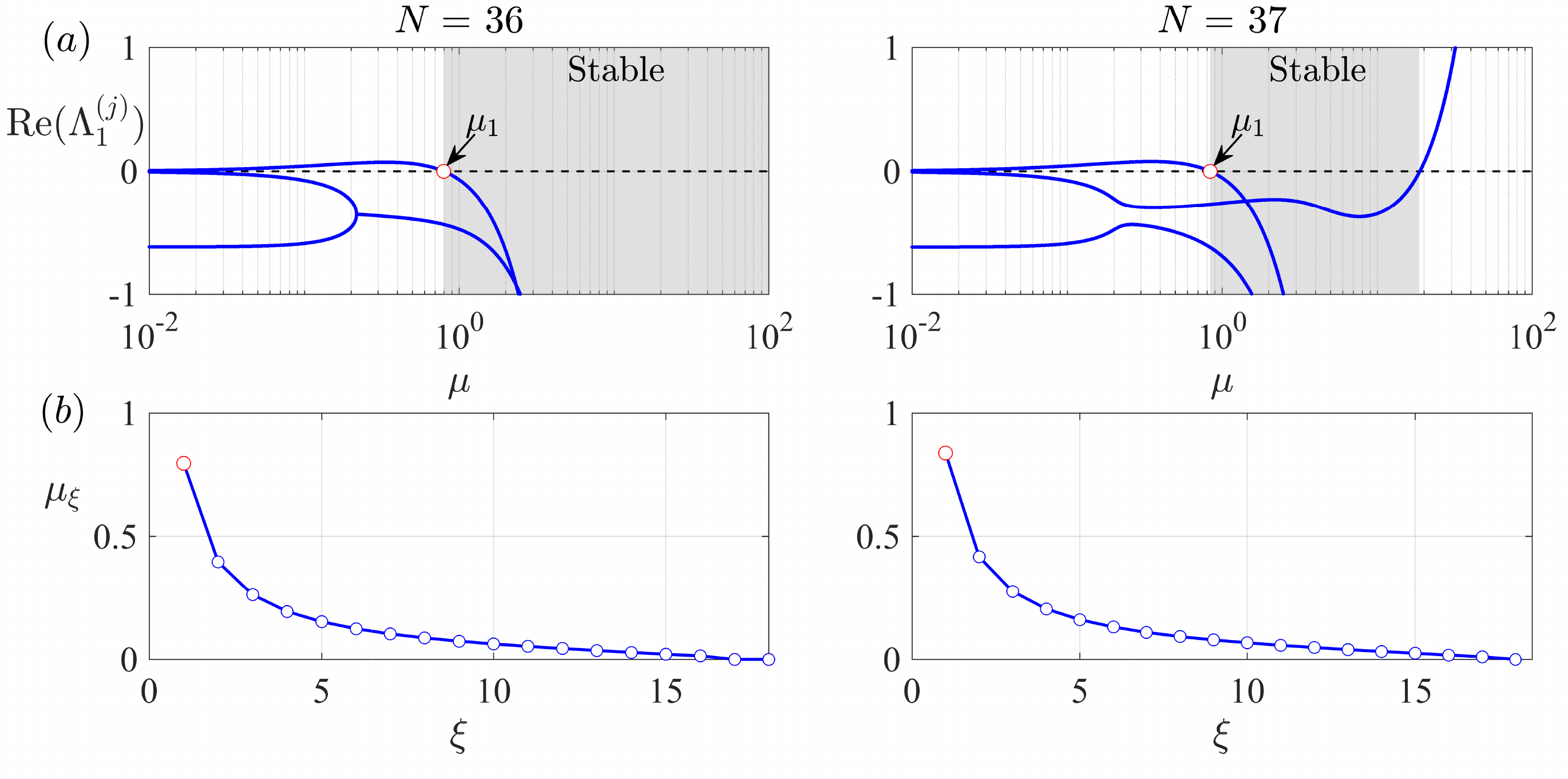}
\end{center}
\caption{
\textcolor{black}{
The linear stability of the periodic state for $N = 36$ (left) and $N = 37$ (right) computed for nearest-neighbor interactions ($p = 1$).
$(a)$ The case $\xi = 1$. The real part of the eigenvalues $\Lambda_1^{(j)}$ for $j = 1, 2, 3$ as a function of $\mu$, where the asymptotic behavior for $\mu \gg 1$ is given in equation \eqref{eq:Lambda_asymp}. The transition from stable (grey shading) to unstable as $\mu$ decreases determines the instability threshold for $\xi  = 1$ (red circle). For $N = 37$, spurious instability is predicted by the amplitude equations \eqref{eqn:dCGLE_intro} for $\mu \gg 1$, a regime inconsistent with the $\mu = O(1)$ assumption.
$(b)$ The instability threshold, $\mu_\xi$, as a function of the wavenumber, $\xi$, for $\xi \leq N/2$. The long-wave mode, $\xi = 1$ (red circle), is the first to destabilize as $\mu$ is decreased.
}
}
\label{fig:MU_instability_plot}
\end{figure} 

\textcolor{black}{The eigenvalues, $\Lambda_\xi^{(j)}$ (for $j = 1, 2, 3$), of $\text{M}_{\xi}(\mu)$ determine the stability of the spatially uniform state \eqref{eqn:dCGLE_periodic_sols}, where the dependence of $\Lambda_\xi^{(j)}$ on $\mu$ is presented in Figure \ref{fig:MU_instability_plot}.}
The onset of instability is determined by the eigenvalue (or one of a pair of complex-conjugate eigenvalues) of $\text{M}_{\xi}(\mu)$ with maximal real part;  we denote this eigenvalue as $\Lambda_\xi(\mu)$ for each wavenumber, $\xi\in \{0,1,\ldots,N - 1\}$. At a given value of $\mu > 0$, the perturbed system \eqref{eqn:eigenvalue_system} is neutrally stable if $\mathrm{Re}(\Lambda_\xi(\mu)) \le 0$ for all $\xi$, and unstable otherwise. We first note, however, that the rotational and temporal invariance of the spatially uniform state \eqref{eqn:dCGLE_periodic_sols} implies that $\Lambda_{0} = 0$ has multiplicity two. The remaining eigenvalue for $\xi = 0$ is $-2\rho^2\text{Re}(\sigma_{2}) = -2\text{Re}(\sigma_{1}) < 0$, corresponding to a stable perturbation from the periodic state for all $\mu > 0$. \textcolor{black}{Hence, if an instability to the periodic state arises, then it is for $\xi\neq 0$, corresponding to the emergence of a nontrivial spatial pattern. Moreover, for $\mu \gg 1$ and $\xi \neq 0$, the diagonal elements of the matrix $\mathrm{M}_\xi(\mu)$ dominate, from which we infer that the eigenvalues of $\mathrm{M}_\xi(\mu)$ are approximated by 
\begin{equation}
\label{eq:Lambda_asymp}
\Lambda_\xi^{(1)} \sim \mu^2 D_2 \Delta_\xi, \quad 
\Lambda_\xi^{(2)} \sim \mu^2(D_1 \Delta_\xi - c_g \nabla_\xi), \quad
\Lambda_\xi^{(3)} \sim\mu^2(D_1^* \Delta_\xi - c_g^* \nabla_\xi),
\end{equation}
corresponding to advection and diffusion of perturbations. 
We recall that $ \mathrm{Re}(D_1) > 0$, $D_2 > 0$, and $\Delta_{\xi} < 0$ for $\xi \neq 0$. Hence, for $c_g = 0$ (arising when $k_c = N/2$), all three eigenvalues have negative real part for $\mu \gg 1$ (as evidenced in Figure \ref{fig:MU_instability_plot} for $N = 36$).
However, for $c_g \neq 0$, there are regimes in which the spatially uniform state is spuriously predicted to be unstable for $\mu \gg 1$ (a regime inconsistent with our $\mu = O(1)$ assumption), as presented in Figure \ref{fig:MU_instability_plot} for $N = 37$.
To sidestep this spurious prediction, we therefore define $\mu_{\xi} > 0$ as the largest value of $\mu$ at which $\text{Re}(\Lambda_{\xi}(\mu_{\xi})) = 0$ and $\text{Re}(\Lambda_{\xi}(\mu))$ is a decreasing function of $\mu$ at $\mu = \mu_\xi$ (see Figure \ref{fig:MU_instability_plot}$(a)$ for reference).
In the $\mu = O(1)$ regime of interest, the system is therefore unstable to perturbations for $\mu < \mu_c = \max_\xi \mu_\xi$.}

\subsection{The onset of spatial modulations}


A crucial feature of the system \eqref{eqn:dCGLE_intro} is the coupling of the complex amplitude, $A_{n}$, with the mean, $B_n$, which acts to drive the instability of the spatially uniform state \eqref{eqn:dCGLE_periodic_sols}.
We observe that variations in $B_n$ act as a source term in the amplitude, $A_n$, thus promoting spatial variations in $A_n$. For large $\mu$, the diffusion term in \eqref{eqn:dCGLE_intro1} counteracts the growth of spatial variations of $A_n$, but this smoothing effect may become subdominant to the source term when $\mu$ is sufficiently small. Likewise, a similar competition between diffusion and the excitation of spatial variations in $B_n$ driven by variations in $A_n$ is apparent in equation \eqref{eqn:dCGLE_intro2}. As a consequence, this coupling provides a positive feedback loop for the emergence of spatial variations, whereas a BFN-like mechanism instead relies on sufficiently small dissipation. In fact, a true BFN instability only arises in the system \eqref{eqn:dCGLE_intro} when the coupling coefficients, $\gamma_i$, and the group speed, $c_g$, vanish. A necessary condition for instability in this artificial case is $\text{Re}(\sigma_2 D^{*}_{1}) < 0$ \cite{ravoux2000stability}.


\textcolor{black}{
By numerically computing the eigenvalues of $\mathrm{M}_\xi(\mu)$, we observe that, similar to the BFN instability, the amplitude equations \eqref{eqn:dCGLE_intro} exhibit a long-wave instability at $\xi = 1$ (see Figure \ref{fig:MU_instability_plot}); hence, in the cases considered here, we have $\mu_c = \mu_1$.
To capture this phenomenon analytically, we consider the special where $N$ is even and $k_c = N/2$, resulting in $c_g = 0$ and $\gamma_3 = 0$. In this case, the spatially uniform state destabilizes \emph{via} a real eigenvalue, for which the corresponding value of $\mu_\xi$ satisfies $\det \mathrm{M}_\xi(\mu_\xi) = 0$. This condition is satisfied by $\mu_\xi = 0$ or when $\mu_\xi$ has a positive solution to
$$ 
\Big[ |D_1|^2 D_2 \Delta_\xi^3 \Big]\mu_\xi^4 
 - 2 \rho^2 \Delta_\xi \Big[D_2 \Delta_\xi\, \mathrm{Re}(\sigma_2 D_1^*) + \nabla_\xi^2\, \mathrm{Re}(\gamma_1^* \gamma_2D_1)\Big]\mu_\xi^2 - 4 \rho^4\nabla_\xi^2\,\mathrm{Im}(\sigma_2\gamma_1^*) \,\mathrm{Im}(\gamma_2) = 0.
$$
As $\nabla_\xi = -\nabla_{N-\xi}$ and $\Delta_\xi$ and $\Delta_{N-\xi}$, the coefficients of this equation are invariant under the mapping $\xi\rightarrow N - \xi$, so $\mu_\xi = \mu_{N - \xi}$ for all $\xi$. Moreover, to elucidate the dependence of $\mu_\xi$ on $\xi$, we consider the limiting case $\xi \alpha \ll 1$. In this limit, we observe that $\nabla_\xi^2 \sim \Delta_\xi \sim -\xi^2$, yielding $\mu_\xi  = \kappa\rho\sqrt{2}/\xi$, where $\kappa$ is a real and positive root of the quartic polynomial 
$$ 
P(\kappa) =  \Big[ |D_1|^2 D_2 \Big]\kappa^4
 +  \Big[D_2\, \mathrm{Re}(\sigma_2 D_1^*) + \mathrm{Re}(\gamma_1^*\gamma_2 D_1)\Big] \kappa^2 - \mathrm{Im}(\sigma_2\gamma_1^*) \,\mathrm{Im}(\gamma_2).
$$
If $P$ has a real and positive root then the form $\mu_\xi \sim \xi^{-1}$ suggests that the first wavenumber to destabilize is $\xi = 1$ (or $\xi = N - 1$, by symmetry); this long-wave instability is thus consistent with our slowly varying approximation of convolutions (see \S\ref{sec:WNL}).
}
\subsection{Dependence on changes to the wave field and lattice parameters}
\label{sec:change_wave}
We proceed to explore the dependence of $\mu_c$ on the inter-droplet spacing, $\delta$, and spatial decay length, $l$, of the wave kernel, $\mathcal{H}$. Motivated by the form of the wave field arising in the bouncing-droplet system \cite{eddi2011information}, a candidate wave kernel that satisfies the assumed periodicity, exponential decay and quasi-monochromaticity may be derived by projecting the form of the dimensionless radially symmetric wave $F(r) = \mathcal{A}J_0(2\pi r)\mathrm{sech}(r/l)$ onto a circle of circumference $2\pi r_0 = N\delta$, where $r_0 = R/\lambda$ \cite{thomson2020hydro}. Here $J_0$ is the Bessel function of the first kind with order zero and $\mathcal{A}$ is the amplitude of the wave. The resultant algebraic form of the wave kernel is
\begin{equation}
\label{eq:wave_kernel_def}
\mathcal{H}(x) = F\bigg(2r_0 \sin\frac{x}{2r_0}\bigg),
\end{equation}
where an example of this wave kernel is given in Figure \ref{fig:RD}$(b)$. For the numerical results presented herein, we consider $\mathcal{A} = 0.1$ \cite{thomson2020hydro}. We note that the qualitative features of these results do not depend on the value of $\mathcal{A}$; increasing $\mathcal{A}$ simply serves to increase the amplitude of the wave field accompanying the equispaced lattice, the main consequence of which is a concomitant decrease in the critical memory, $M_c$.

\begin{figure}[h!]
\begin{center}
\includegraphics[width=\textwidth]{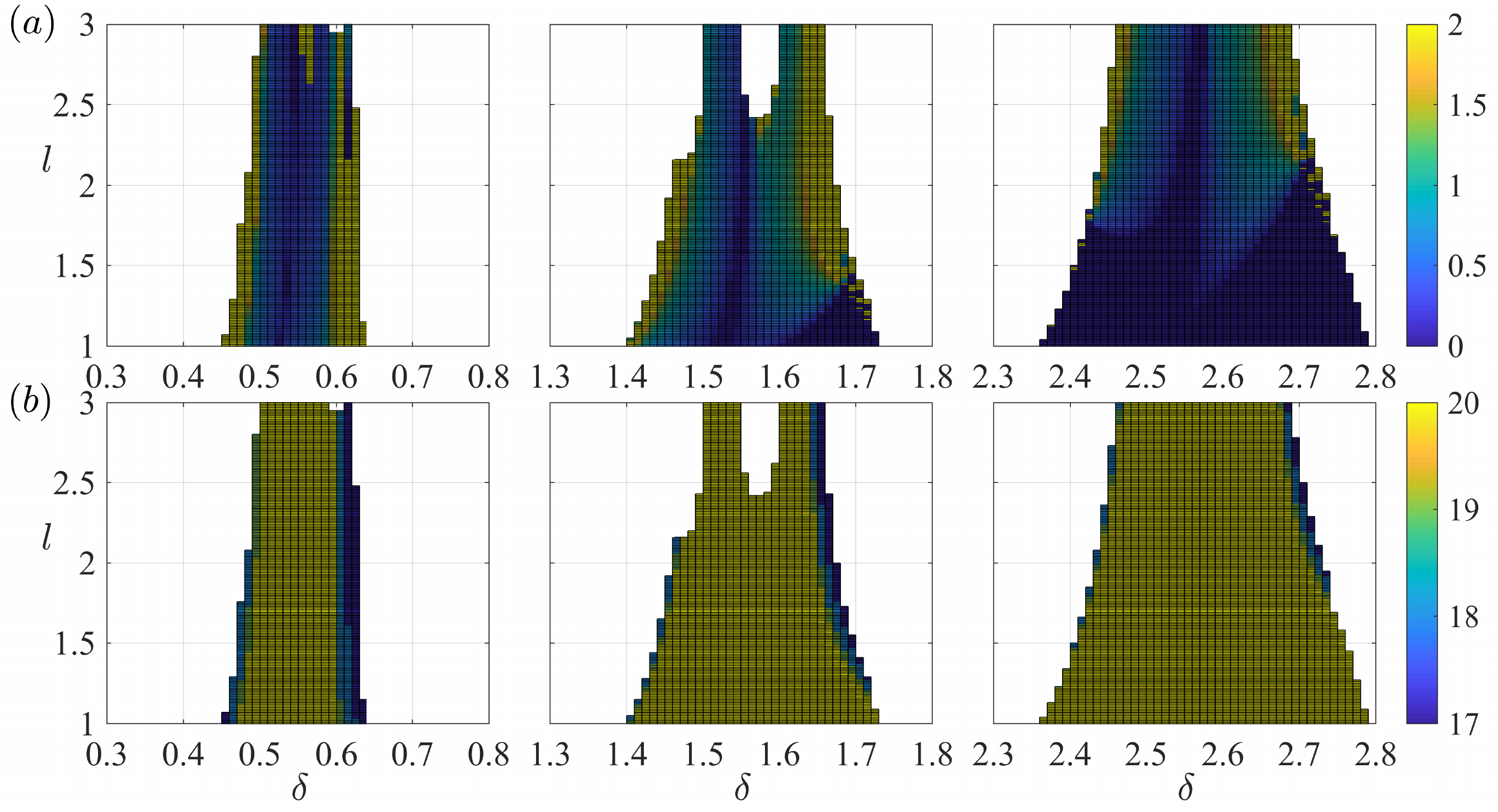}
\end{center}
\caption{The onset of spatial modulations for $N = 40$ droplets, as predicted by the linear stability analysis of the amplitude equations \eqref{eqn:dCGLE_intro} for the wave kernel defined in equation \eqref{eq:wave_kernel_def} and $p = 1$ nearest neighbors.
$(a)$ When the initial instability of the equidistant lattice arises \emph{via} a supercritical Hopf bifurcation, we color-code each value of the spacing parameter, $\delta$, and the decay length, $l$, by $\mathrm{min}(\mu_c, 2)$, where spatial instabilities arise for $\mu < \mu_c$. Our theory is valid when $\mu_c = O(1)$. When $\mu_c = 0$, the periodic state is predicted to be unconditionally stable. Large values of $\mu_c$ (i.e.\ those exceeding the threshold of 2) arise near the boundary between super- and sub-critical Hopf bifurcations.
$(b)$ The corresponding value of $k_c$, as predicted by the linear stability analysis, demonstrating the correlation between large $\mu_c$ and $k_c < N/2$.
}
\label{fig:Mu_c}
\end{figure}

As presented in Figure \ref{fig:Mu_c}, the dependence of the onset of spatial modulations on the system parameters, $l$ and $\delta$, can be quite intricate. Near the boundaries between super- and sub-critical Hopf bifurcations (where geometric and subcritical Hopf instabilities arise in the white regions in Figure \ref{fig:Mu_c}), $\mu_c$ can be very large---a feature that appears to be correlated with $k_c$ departing from $N/2$---inconsistent with the $\mu = O(1)$ assumption under which the amplitude equations \eqref{eqn:dCGLE_intro} were derived. Away from these boundaries, we observe regions in which $\mu_c = O(1)$; indeed, simulation of the lattice system \eqref{eqn:drops_nondim} reveals favorable agreement of the numerical and theoretical instability threshold. (This agreement improves as $\alpha = 2\pi/N$ becomes smaller, consistent with our assumptions; see supplementary material.) Near the middle of each `band' in which supercritical Hopf bifurcations arise, we observe that $\mu_c = 0$, corresponding to the prediction of unconditional stability of the spatially uniform solution \eqref{eqn:dCGLE_periodic_sols}. Finally, we remark that $\mu_c = O(1)$ is most apparent for $l \sim \delta$, a regime in which the inter-droplet coupling is dominated by nearest-neighbor interactions, as might be anticipated from our local approximation of convolutions (see \S \ref{sec:WNL}). When $l$ is much smaller than $\delta$, each droplet only interacts weakly with all other droplets (including its nearest neighbors), and spatial variations are less propitious.

\section{Numerical solutions}
\label{sec:numerical_solns}

We proceed to explore the nonlinear dynamics predicted by the amplitude equations \eqref{eqn:dCGLE_intro} beyond the onset of spatial variations, $\mu < \mu_c$. The amplitude equations are evolved using a spectral method over the droplet number, $n$, and a fourth-order Runge-Kutta method in time, for which the linear terms are transformed using an integrating factor (see Appendix \ref{app:numerical} for details) \cite{milewski1999pseudospectral}. Initially considering $\mu$ just below the instability threshold, $0 < \mu_c - \mu \ll 1$, we evolve the amplitude equations \eqref{eqn:dCGLE_intro} from the initial condition $A_n = \rho + \zeta\sin(n \alpha)$ and $B_n = \zeta\sin(n \alpha)$ ($\zeta\ll 1$) until a periodic state is attained. Thereafter, we decrement $\mu$ by 0.02 and initialize the following simulation at the final values obtained in the preceding simulation. The MATLAB code used to simulate these dynamical states is provided in the supplementary material.

As suggested by Figure \ref{fig:Mu_c}, there is a vast parameter space we could explore with equations \eqref{eqn:dCGLE_intro} by varying the parameters $l$, $\delta$, and $N$. To fix ideas, we focus on the case of two adjacent droplet numbers, $N =40$ and $N = 41$, and set $l = \delta = 2.6$, which serves to elucidate the key phenomenology exhibited by the amplitude equations \eqref{eqn:dCGLE_intro}. A deeper exploration of the parameter space is reserved for future work. Before we present the solutions, we note that the only variable parameter in the amplitude equations \eqref{eqn:dCGLE_intro} is $\mu$, since the coefficients are fixed for a particular choice of wave kernel \eqref{eq:wave_kernel_def} and constituent parameters $l$, $\delta$, and $N$. Thus, varying $\mu$ corresponds to traversing a \emph{particular path through parameter space}, in contrast to varying each coefficient in the amplitude equations independently. As we shall see, this variation in $\mu$ gives rise to a series of bifurcations between qualitatively different dynamical behaviors.

\begin{figure}[h!]
\begin{center}
\includegraphics[width=\textwidth]{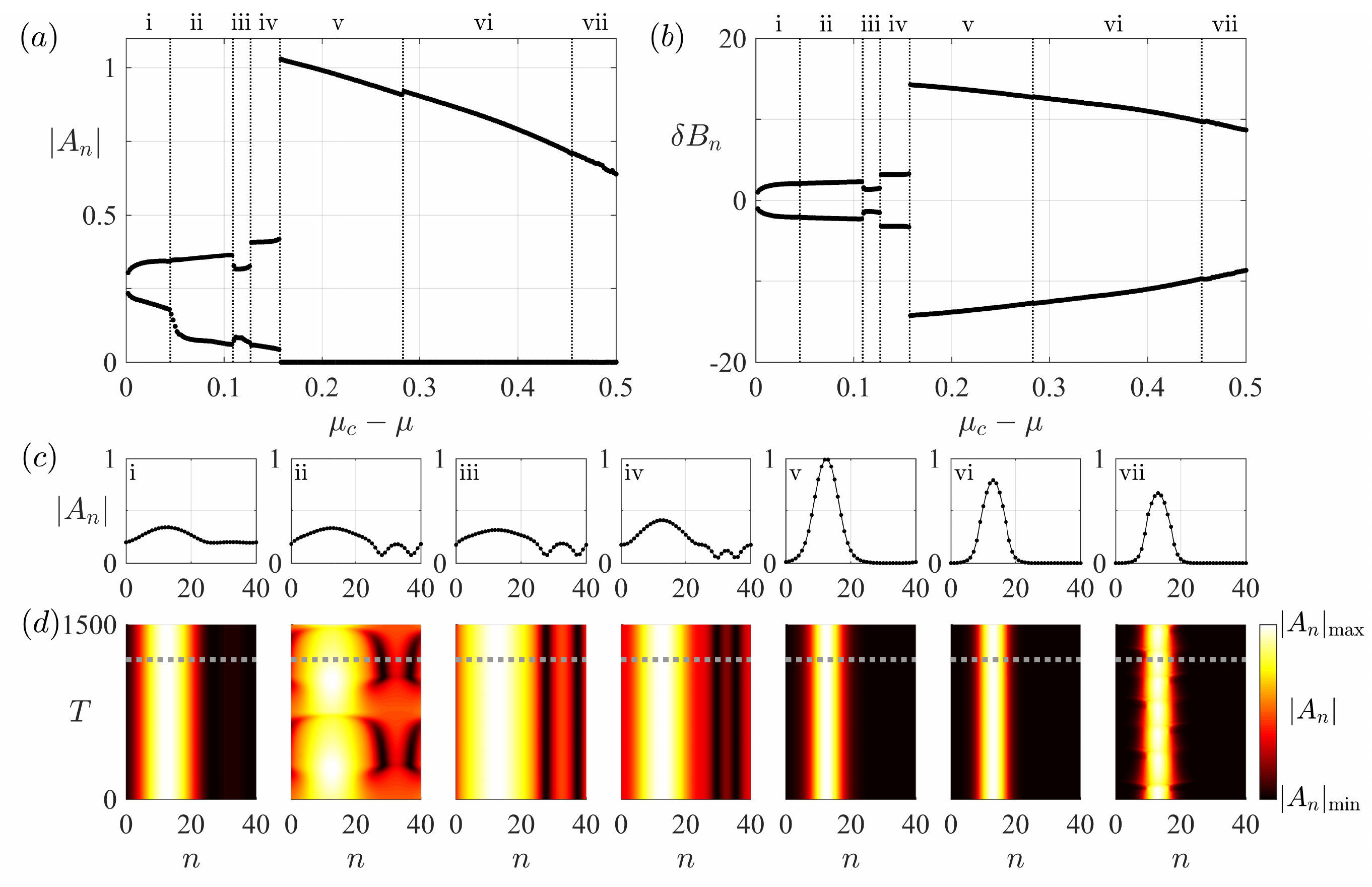}
\end{center}
\caption{
The nonlinear dynamics predicted by the amplitude equations \eqref{eqn:dCGLE_intro} for $N = 40$ droplets and $\mu < \mu_c = 0.851$. 
$(a)$ The upper and lower bounds of $|A_n(T)|$ attained over the entire simulation. We identify seven dynamical regimes, which are denoted by Roman numerals and divided by the dotted vertical lines.
$(b)$ The upper and lower bounds of $\delta B_n(T) = B_n(T) - \langle B_n(T)\rangle$ attained over the entire simulation, where angled brackets denote the average over $n$.
$(c)$ Snapshots of $|A_n|$ ($T$ fixed) for each dynamical regime, where $\mu_c - \mu$ takes values 
(i) 0.03,
(ii) 0.08,
(iii) 0.12,
(iv) 0.14,
(v) 0.2,
(vi) 0.4, and
(vii) 0.48.
$(d)$ Corresponding spacetime plots of $|A_n(T)|$ for the parameter values in $(c)$. The dashed lines correspond to the snapshot time, $T = 1200$.
The system parameters here $\delta = 2.6$, $l = 2.6$, and $p = 1$ nearest neighbors, yielding $k_c = 20$ and $\nu_c = 1.585$.
}
\label{fig:N40Simulations}
\end{figure}

\begin{figure}[h!]
\begin{center}
\includegraphics[width=\textwidth]{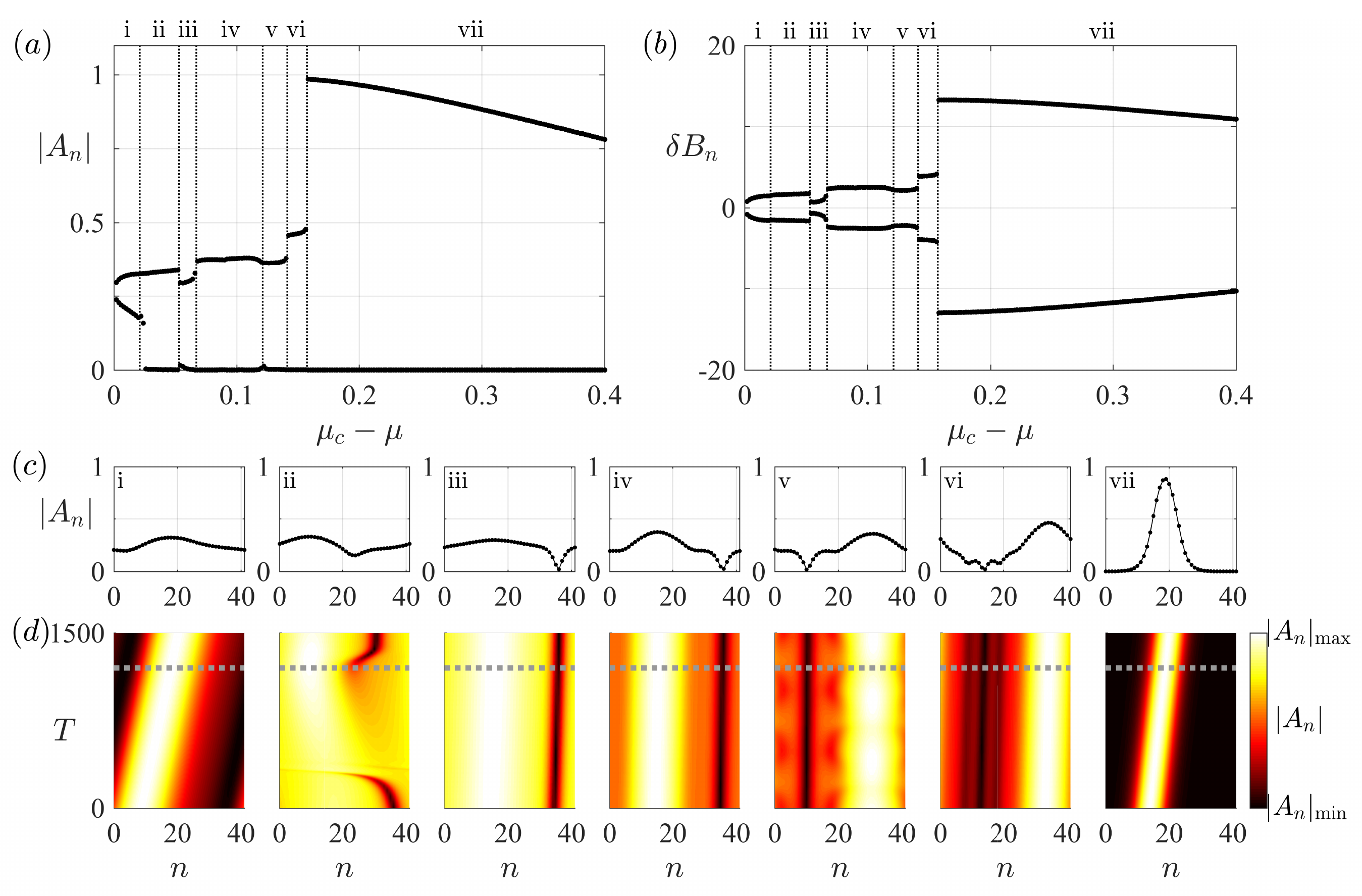}
\end{center}
\caption{
The nonlinear dynamics predicted by the amplitude equations \eqref{eqn:dCGLE_intro} for $N = 41$ droplets and $\mu < \mu_c = 0.852$. 
$(a)$ The upper and lower bounds of $|A_n(T)|$ attained over the entire simulation. We identify seven dynamical regimes, which are denoted by Roman numerals and divided by the dotted vertical lines.
$(b)$ The upper and lower bounds of $\delta B_n(T) = B_n(T) - \langle B_n(T)\rangle$ attained over the entire simulation, where angled brackets denote the average over $n$.
$(c)$ Snapshots of $|A_n|$ ($T$ fixed) for each dynamical regime, where $\mu_c - \mu$ takes values 
(i) 0.014,
(ii) 0.04,
(iii) 0.06,
(iv) 0.08,
(v) 0.1,
(vi) 0.15, and
(vii) 0.3.
$(d)$ Corresponding spacetime plots of $|A_n(T)|$ for the parameter values in $(c)$. The dashed lines correspond to the snapshot time, $T = 1200$.
The system parameters here $\delta = 2.6$, $l = 2.6$, and $p = 1$ nearest neighbors, yielding $k_c = 20$ and $\nu_c = 1.582$.
}
\label{fig:N41Simulations}
\end{figure}

In the case of $40$ droplets (see Figure \ref{fig:N40Simulations}), close to the onset of spatial modulations at $\mu = \mu_c$, we first observe a time-independent solution for $|A_n|$ (panel (i)), which destabilizes into a periodic, breather-like state (panel (ii)) with dips apparent in $|A_n|$. These dips persist as $\mu_c - \mu$ is increased (panels (iii) and (iv)), but $|A_n|$ is instead constant in time. When $\mu_c - \mu$ is sufficiently large (panels (v) and beyond), a robust stationary bright soliton emerges over a large range of $\mu$, before destabilizing far from the instability threshold (panel (vii)), leading to chaotic fluctuations in the soliton form. We note that when $N =40$, the group speed, $c_g$, is identically zero. For $N = 41$ droplets and non-zero group speed (Figure \ref{fig:N41Simulations}), similar dynamical transitions occur, although the asymmetry of the system instead yields traveling waves and propagating solitons (panels (i) and (vii)). Notably, we also observe parameter regimes for which dark breathers (panel (ii)) and dark solitons (panels (iii)--(v)) arise, characterized by the sharp dips in the amplitude, $|A_n(T)|$, towards zero, which are not present when $N = 40$. All of the aforementioned features arise over a large range of $N$, and thus appear to be canonical features of the discrete amplitude equations \eqref{eqn:dCGLE_intro}. We also note that the jumps in the bounds of $|A_n|$ and $B_n$ are indicative of hysteresis between dynamical states, an effect to be explore in greater detail elsewhere.

\section{Discussion \& conclusion}
\label{sec:concl}
In this paper, we have presented a rigorous mathematical framework to derive a discrete set of amplitude equations from a driven and dissipative oscillator model, inspired by the physics of droplet lattices bouncing on a vibrating fluid bath. Our systematic derivation provides a direct link between the constitutive properties of the lattice model \eqref{eqn:drops_nondim} (specifically, the wave kernel, $\mathcal{H}$) and the coefficients arising in the amplitude equations \eqref{eqn:dCGLE_intro}. A linear stability of the amplitude equations \eqref{eqn:dCGLE_intro} reveals the importance of the coupling to the discrete mean equation \eqref{eqn:dCGLE_intro2} in destabilising the system from a spatially uniform state, leading to spatial modulations in the droplet amplitude following a second bifurcation, similar in spirit to the BFN instability. Beyond this second bifurcation, numerical solutions of the amplitude equations \eqref{eqn:dCGLE_intro} reveal a fascinating family of dynamical behaviours including bright and dark solitons, breather states, and traveling waves. As computational models of the droplet system advance \cite{durey2020faraday}, the predictions of the amplitude equations may be compared against direct numerical simulations of the droplet dynamics, paving the way for further experimental investigation \cite{thomson2020collective}. In particular, a tantalizing prospect is to hunt for the emergence of so-called chimera states \cite{abrams2004chimera,sethia2008clustered}, thought to be ubiquitous in coupled oscillators subject to non-local coupling, but have been shown to exist in only a handful of experimental systems to date \cite{nkomo2013chimera,martens2013chimera,wojewoda2016smallest,totz2018spiral}. 

We conclude with the proposition that the framework used to derive the amplitude equations \eqref{eqn:dCGLE_intro} applies to a more general class of oscillator model of canonical form
\begin{subequations}
\label{eqn:dynamic_oscillator}
\begin{align}
\ddot{x}_{n} + \dot{x}_{n} &= -\frac{\partial h}{\partial x}(x_{n},t),\\
\mathcal{P} h &= \sum_{m=1}^{N} \mathcal{H}(x - x_m),
\end{align}
\end{subequations}
with the linear operator $\mathcal{P} = \partial/\partial t + \nu$ in equation \eqref{eqn:drops_nondim_2} serving as a particular example. The novelty of the model \eqref{eqn:dynamic_oscillator} is that the inter-particle coupling potential, $h$, is \emph{dynamic}, continuously evolving with the particle motion, rather than being fixed in space or with respect to the particles. Other potential choices of $\mathcal{P}$ are numerous, and could lead to an even richer family of dynamics. For a particular choice of $\mathcal{P}$, if the bifurcation leading to instability of the oscillator is of supercritical Hopf type, then one should expect a complex Ginzburg-Landau equation in the vicinity of the bifurcation point. However, the precise form of the amplitude equations will change depending on the type of primary bifurcation that arises and the inherent symmetries of the system \cite{aranson2002world,cross1993pattern}. Such an investigation may lead to further insights into the dynamics and pattern-forming behaviour of active particles in complex environments \cite{bechinger2016active}.

\begin{appendices}

\section{Derivation of the Ginzburg-Landau and mean equations}
\label{sec:sl_derive}

Further details are provided of the multiple-scales expansion leading to the complex Ginzburg-Landau and mean equations \eqref{eqn:dCGLE_intro}. The general procedure toward obtaining equations \eqref{eqn:dCGLE_intro} is to substitute the asymptotic expansions \eqref{eqn:ms_expansions} into \eqref{eqn:drops_nondim} and gather successive powers of $\varepsilon$. At each successive order, we suppress resonant terms proportional to $\text{e}^{\myi \phi_{n}(t)}$ (where $\phi_{n}(t) = k_{c}n\alpha + \omega_{c}t$) or those constant in $t$. To extract all relevant terms at each order, we must introduce auxiliary variables to solve for the free surface, $h$, and also expand convolutions in the manner summarized in \S\ref{sec:WNL}. For notational efficiency, we denote $\mathcal{H}_m = \mathcal{H}(m\delta)$, $\mathcal{H}_m' = \mathcal{H}'(m\delta)$, and so forth.

At leading order we obtain
\begin{equation}
\label{eqn:wn1}
\pd{\hzero}{x}\bigg\vert_{x = n \delta} = 0,\quad \hzero(x) = \frac{1}{\nu_{c}}\sum_{m = 1}^{N}\Hkern(x - m\delta),
\end{equation}
reflecting the fact that the free-surface gradient beneath each droplet vanishes in the steady-state. Consequently, all odd-derivatives of $\hzero$ vanish beneath each droplet at equilibrium, a fact we will make repeated use of in simplifying the forthcoming terms in arising our expansion.


The problem at $O(\varepsilon)$ has already been discussed in \S\ref{sec:WNL} and thus we proceed directly to $O(\varepsilon^2)$, remembering that terms from \eqref{eqn:disp_relation_expand} are promoted to $O(\varepsilon^3)$ after setting $\alpha = \mu\varepsilon$. The equations at $O(\varepsilon^2)$ are
\begin{align}
\label{eqn:xn2}
\pdd{2}{\xtwo_n}{t} + \pd{\xtwo_n}{t} &= -\left\{\pd{\htwo}{x} + \xtwo_{n}\frac{\partial^2 h^{(0)}}{\partial x^2} + \xone_{n}\frac{\partial^2\hone}{\partial x^2}\right\}\Big\vert_{x = n \delta},\\
\label{eqn:h2}
\frac{\partial\htwo}{\partial t} + \nu_{c}\htwo &= \hzero+
\sum_{m=1}^{N}\left\{\frac{1}{2}x^{(1)2}_{m}\Hkern''(x - m\delta)-\xtwo_{m} \Hkern'(x - m\delta)\right\}.
\end{align}
Following the procedure outlined in \S\ref{sec:solve_wf}, we solve for $h^{(2)}$ by introducing two further auxiliary variables, $Y_{n}$ and $Z_n$, satisfying
\begin{equation}
\label{eqn:X2_X3}
\pd{Y_{n}}{t} + \nu_{c}Y_{n} = \frac{1}{2}x^{(1)2}_{n} ,\qquad 
\pd{Z_{n}}{t} + \nu_{c}Z_{n} = \xtwo_{n},
\end{equation}
chosen to match the coefficients of the wave kernel on the right-hand side of \eqref{eqn:h2}. Hence, a particular solution of \eqref{eqn:h2} is
\begin{equation}
\label{eqn:h2_sol}
\htwo = \frac{1}{\nu_{c}}\hzero +\sum_{m=1}^N\left\{Y_{m}\Hkern''(x - m\delta) - Z_{m}\Hkern'(x - m\delta)\right\}.
\end{equation}
By substituting the form of $\xone_n$, given by equation \eqref{eqn:xone_ansatz}, into the first equation of \eqref{eqn:X2_X3}, we find
\begin{equation}
\label{eqn:Whyn}
Y_n = \frac{1}{\nu_{c}}\left[|A_n|^2 + \frac{1}{2}B_n^2\right] + B_n\left[\frac{A_n}{\nu_{c} + \myi\omega_{c}}\mye^{\myi\phi_{n}}+\text{c.c.}\right] + \frac{1}{2}\left[\frac{A_n^2}{\nu_{c} + 2\myi\omega_{c}}\mye^{2\myi\phi_{n}} + \text{c.c.}\right].
\end{equation}
Substituting \eqref{eqn:h2_sol}--\eqref{eqn:Whyn} into \eqref{eqn:xn2}, and then using \eqref{eqn:wn1}, yields
\begin{multline}
\label{eqn:L_x2}
\mathcal{L}_{n}{\bm{x}}^{(2)} = \Big[ B_n + \{A_n \mye^{\myi \phi_n} + \cc \} \Big]
\bigg[\frac{1}{\nu_c}\sum_{m = 1}^N B_{n-m}\Hkern'''_m 
+
\bigg\{\mye^{\myi \phi_n} \sum_{m = 1}^N \frac{A_{n-m}\mye^{-\myi k_c m \alpha } } {\nu_c + \myi \omega_c}\Hkern'''_m + \cc
\bigg\}
\bigg] \\
- \frac{1}{\nu_c}\sum_{m = 1}^N \bigg[\frac{1}{2}B_{n-m}^2 + |A_{n-m}|^2\bigg]\Hkern'''_m
- \bigg[ \mye^{\myi\phi_n}\sum_{m = 1}^N \frac{A_{n-m}B_{n-m} }{\nu_c + \myi \omega_c }\mye^{-\myi k_c m \alpha } \Hkern'''_m + \cc
\bigg] \\
-\frac{1}{2}\bigg[ \mye^{2 \myi \phi_n}\sum_{m = 1}^N \frac{A_{n-m}^2 \mye^{- 2\myi k_c m \alpha }  \Hkern'''_m }{\nu_c + 2 \myi \omega_c } + \cc 
\bigg].
\end{multline}
Analogous to \S\ref{sec:approx_convs}, we now use the slowly varying approximation \eqref{eqn:F_expansion} to expand $B_{n-m}$ on the right-hand side of \eqref{eqn:L_x2}, and similarly for all other terms of this form (such as $A_{n-m}$, $B_{n-m}^2$, etc.).
After some arduous algebra, we reduce \eqref{eqn:L_x2} to the highly simplified form 
\begin{multline}
\label{eqn:L_x2_b}
\mathcal{L}_{n}{\bm{x}}^{(2)} =
2\mathrm{Re}\bigg[ \sum_{m=1}^N \frac{\mye^{-\myi k_c m \alpha } \Hkern_m''' }{\nu_c + \myi\omega_c }   \bigg] |A_n|^2
 +
\alpha\big[  \hat{\gamma}_1 \mye^{\myi \phi_n} A_n \nabla B_n + \cc
\big] +
2\alpha \mathrm{Re}\big[\hat{\gamma}_2 A_n\nabla A_n^*  
\big] \\
+ \big[ \hat{c}_1 A_n^2 \mye^{2\myi \phi_n} + \cc \big]
+ \alpha\big[\hat{c}_2 \mye^{2\myi \phi_n} A_n \nabla A_n  + \cc
\big] + O(\alpha^2),
\end{multline}
where the complex coefficients are defined as
\begin{multline}
\label{eqn:comp_coeffs_1}
\hat{\gamma}_1 =   \sum_{m = 1}^N \bigg( \frac{\mye^{-\myi k_c m \alpha } }{\nu_c + \myi\omega_c} 
- \frac{1}{\nu_c}\bigg) a_m \Hkern'''_m, \qquad\hat{\gamma}_2 =  \sum_{m = 1}^N\bigg(\frac{1}{\nu_c} - \frac{ \mye^{\myi k_c m \alpha }}{\nu_c - \myi\omega_c } \bigg) a_m \Hkern'''_m,\\
\hat{c}_1 = \sum_{m=1}^N \bigg[ \frac{\mye^{-\myi k_c m \alpha }  }{\nu_c + \myi \omega_c } -  \frac{\mye^{-2\myi k_c m \alpha } }{2( \nu_c + 2\myi\omega_c ) }\bigg]\Hkern'''_m, \quad 
\hat{c}_2 =
\sum_{m=1}^N \bigg(\frac{\mye^{-2\myi k_c m \alpha }}{ \nu_c + 2\myi \omega_c }  - \frac{ \mye^{-\myi k_c m \alpha } }{\nu_c + \myi \omega_c }\bigg) a_m \Hkern_m'''.
\end{multline}

To apply the weak-asymmetry approximation developed in \S\ref{sec:weak_asymm} to the coefficient of $|A_n|^2$, we first note that
$$2\mathrm{Re}\bigg[ \sum_{m=1}^N \frac{\mye^{-\myi k_c m \alpha } \Hkern_m''' }{\nu_c + \myi\omega_c }   \bigg]  = 2\mathrm{Re}\bigg[\frac{\myi}{\nu_c + \myi\omega_c }\bigg] \sum_{m=1}^N (-1)^m \sin(m\alpha\chi) \Hkern_m''', $$ 
where $\chi = N/2 - k_c$.
By a similar argument to which the group velocity was promoted to $O(\varepsilon^3)$ in \S\ref{sec:weak_asymm}, the coefficient of $|A_n|^2$ has size $O(\alpha)$ for $\alpha \ll 1$ and $\chi = O(1)$, so the corresponding term in \eqref{eqn:L_x2_b} should likewise appear at $O(\varepsilon^3)$. We thus write
\begin{equation*}
2\mathrm{Re}\bigg[ \sum_{m=1}^N \frac{\mye^{-\myi k_c m \alpha } \Hkern_m''' }{\nu_c + \myi\omega_c }   \bigg] |A_n|^2 = \alpha \hat{\gamma}_3 |A_n|^2,
\end{equation*}
where the $O(1)$ real coefficient $\hat{\gamma}_3$ is defined as
\begin{equation}
\label{eqn:comp_coeffs_2}
\hat{\gamma}_3 = \frac{2}{\alpha}\mathrm{Re}\bigg[ \sum_{m=1}^N \frac{\mye^{-\myi k_c m \alpha } \Hkern_m''' }{\nu_c + \myi\omega_c }   \bigg].
\end{equation}

In a similar spirit, we deduce that the coefficient $\hat{c}_{1}$ is size $O(\alpha)$ when $\chi \neq 0$ and zero otherwise. Therefore the non-secular terms in \eqref{eqn:L_x2_b} (those that are proportional to $\text{e}^{2\myi\phi_{n}}$) are both of size $O(\alpha)$ and so should actually appear at $O(\varepsilon^3)$. However, as these terms will still be non-secular at that order, they play no role in the derived amplitude equations for $A_n$ and $B_n$. We conclude that all the inhomogeneities in \eqref{eqn:L_x2_b} (which are of size $O(\alpha)$) should instead appear at $O(\varepsilon^3)$. Hence, at $O(\varepsilon^2)$, we have $\mathcal{L}_{n}{\bm{x}}^{(2)} =  0$, which is identical to the problem for ${\bm{x}}^{(1)}$.
Akin to the solution \emph{ansatz} at $O(\varepsilon)$, we therefore pose
$$\xtwo_{n} = E_n(T) + \big[C_n(T)\mye^{\myi\phi_{n}} + \text{c.c.}\big],\quad
Z_{n} = \frac{1}{\nu_{c}}E_n(T) + \bigg[\frac{C_n(T)}{\nu_{c} + \myi\omega_{c}}\mye^{\myi\phi_{n}} + \text{c.c.}\bigg], $$
which, when applying the approximation to the convolution (see \S\ref{sec:approx_convs}), satisfies the inhomogeneous problem to leading order. The $O(\alpha^2)$ terms that arise out of the approximation to this convolution appear at $O(\varepsilon^4)$, which is beyond the order presented in this calculation.


At $O(\varepsilon^3)$, we have a system for $\xthree_n$ and $\hthree$, namely
\begin{multline}
\label{eqn:xthree}
\pdd{2}{\xthree_{n}}{t} + \pd{\xthree_{n}}{t} + \xthree_{n}\frac{\partial^2 \hzero}{\partial x^2}\bigg\vert_{x = n \delta} = -\left[2\frac{\partial^2\xone_{n}}{\partial t\partial T} + \pd{\xone_{n}}{T}\right]\\ - \left[\frac{\partial h^{(3)}}{\partial x} + \xone_{n}\frac{\partial^2 \htwo}{\partial x^2} + \frac{1}{2}x^{(1)2}_{n}\frac{\partial^3 \hone}{\partial x^3} + \xtwo_{n}\frac{\partial^2 \hone}{\partial x^2} + \frac{1}{6}x^{(1)3}_{n}\frac{\partial^4\hzero}{\partial x^4}\right]\Big\vert_{x = n \delta}
\end{multline}
and
\begin{multline}
\label{eqn:hthree}
\pd{\hthree}{t} + \nu_{c}\hthree = -\left[\pd{\hone}{T} - \hone\right] \\+ \sum_{m=1}^N\bigg\{\xone_{m}\xtwo_{m}\Hkern''(x - m \delta) - \xthree_{m}\Hkern'(x - m \delta) - \frac{1}{6}x^{(1)3}_{m}\Hkern'''(x - m \delta)\bigg\}.
\end{multline}
Appended to the right-hand side of \eqref{eqn:xthree} will be the terms promoted from both $O(\varepsilon)$ and $O(\varepsilon^2)$. We follow an identical procedure to our analysis at $O(\varepsilon)$ and $O(\varepsilon^2)$. First we introduce three auxiliary variables (\S\ref{sec:solve_wf}), one for each of the three inhomogeneities in \eqref{eqn:hthree}, and then solve for $\hthree$. We then substitute this solution into \eqref{eqn:xthree}, along with the droplet positions and wave field terms computed from lower orders, and then apply the slowly varying approximation (\S\ref{sec:approx_convs}) to reduce discrete convolutions to spatially local expressions. This procedure rise to a system of the form $\mathcal{L}_n{\bm{x}}^{(3)} = \mbox{RHS}$, where the right-hand-side (RHS) is composed of terms that are constant in $t$, terms with coefficients $\mye^{\pm \myi \phi_n(t)}$, and non-secular terms (whose form can be ignored at this stage). For a bounded solution, we require that the constant and $\mye^{\myi\phi_n}$ secular terms have vanishing coefficients, which yields the following evolution equations for the complex amplitude, $A_n$, and the real drift, $B_n$:
\begin{subequations}
\label{eqn:GLapp}
\begin{align}
\label{eqn:sll}
\hat{\sigma}_0\sdone{A_n}{T} {+ \mu^2 \hat{c}_g \nabla A_n} &= \hat{\sigma}_1 A_n - \hat{\sigma}_2 |A_n|^2 A_n + \mu \hat{\gamma}_1 A_n \nabla B_n +  \mu^2 \hat{D}_1 \Delta A_n, \\
\label{eqn:oedrift}
\hat{b}_0\sdone{B_n}{T} &= \mu^2\hat{D}_2\Delta B_n + 2\mu \mathrm{Re}\big[\hat{\gamma}_2 A_n\nabla A_n^*\big]  {+ \mu \hat{\gamma}_3 |A_n|^2}
,
\end{align}
\end{subequations}
where $\mu = \alpha/\varepsilon$ and we have neglected terms whose coefficients are of size $O(\alpha)$ at $O(\varepsilon^3)$ (\S\ref{sec:weak_asymm}).
We note that we cannot determine the higher-order corrections ($C_n$ and $E_n$) without proceeding to $O(\varepsilon^4)$ and higher. However, a satisfactory approximation is obtained by considering $A_n$ and $B_n$ alone, which form a closed system.

Recalling that $\mathcal{D}_k(\lambda; \nu)$ is the dispersion relation \eqref{eqn:disp_relation}, the coefficients (other than the $\hat{\gamma}_i$ defined in equations \eqref{eqn:comp_coeffs_1} and \eqref{eqn:comp_coeffs_2}) appearing in \eqref{eqn:sll} are as follows:
$$
\hat{\sigma}_0 = \pd{\mathcal{D}_{k_c}}{\lambda}(\myi\omega_c; \nu_c),\quad 
\hat{\sigma}_1 = \pd{\mathcal{D}_{k_c}}{\nu}(\myi\omega_c; \nu_c),
\quad
$$
$$
\hat{\sigma}_2 = \frac{3}{2\nu_c}\sum_{m=1}^N \Hkern''''_m - \sum_{m=1}^N\Hkern''''_m \mathrm{Re}\bigg[\frac{\mye^{-\myi k_c m \alpha }}{\nu_c + \myi \omega_c}\bigg] + 
\frac{1}{2}\sum_{m=1}^N \frac{\mye^{-2\myi k_c m \alpha } \Hkern''''_m }{\nu_c + 2\myi\omega_c}-
\sum_{m=1}^N \frac{\Hkern''''_m \mye^{-\myi k_c m \alpha }}{\nu_c + \myi\omega_c},
$$

$$
\hat{D}_1 = \frac{1}{\nu_c + \myi\omega_c}\sum_{m=1}^{N} b_m \mye^{-\myi k_c m \alpha} \Hkern''_m,
$$
while those that appear in \eqref{eqn:oedrift} are
$$
\hat{b}_0 = \pd{\mathcal{D}_0}{\lambda}(0;\nu_c) =   1 + \frac{1}{\nu_c^2}\sum_{m = 1}^N \Hkern''_m,\qquad \hat{D}_2 = \frac{1}{\nu_c}\sum_{m=1}^{N} b_m \Hkern''_m.
$$
Upon dividing \eqref{eqn:sll} by $\hat{a}_{0}$ and \eqref{eqn:oedrift} by $\hat{b}_{0}$ we arrive at equations \eqref{eqn:dCGLE_intro} in the main text, where $(c_g,\sigma_1,\sigma_2,\gamma_1,D_1) = (\hat{c}_g,
\hat{\sigma}_1,\hat{\sigma}_2,\hat{\gamma}_1,\hat{D}_1)/\hat{\sigma}_0$ and $(D_2,\gamma_2,\gamma_3) = (\hat{D}_2,\hat{\gamma}_2,\hat{\gamma}_3)/\hat{b}_0$.

\section{Approximation of convolutions}
\label{sec:conv_approx}

We introduce an interpolating polynomial of degree $2p$ passing through the points $F_{n-p}, \ldots, F_{n+p}$, where $F_m = F(m\alpha)$, so that 
\begin{equation}
\label{eq:poly_exp_exact}
F_{n-m} = F_n - m \mathscr{D}_1 F_n + \frac{m^2}{2}\mathscr{D}_2 F_n - \ldots + \frac{m^{2p}}{(2p)!}\mathscr{D}_{2p}F_n
\end{equation}
for $|m| \leq p$. 
The operators $\mathscr{D}_j/\alpha^j$ are the symmetric finite difference operators approximating the $j^\mathrm{th}$ derivative of $F(\theta)$ over $2p + 1$ grid points spaced $\alpha$ apart. Hence, as $F^{(j)}(\alpha n) = O(1)$ (by assumption) and
$$\frac{\mathscr{D}_jF_n}{\alpha^j} = \frac{\mathrm{d}^jF}{\mathrm{d}\theta^j}( n\alpha) + O(\alpha^{q})$$ 
for some integer $q > 0$, we conclude that $\mathscr{D}_jF_n/\alpha^j = O(1)$.
Therefore, for $m = -p, \ldots, p$, we may recast \eqref{eq:poly_exp_exact} in the following form:
\begin{equation*}
F_{n-m} = F_n - \alpha m \nabla F_n + \alpha^2\frac{m^2}{2}\Delta F_n  + O(\alpha^3),
\end{equation*}
where $\nabla$ and $\Delta$ are the central finite difference operators approximating the first and second derivatives of $F(\theta)$ using $2p + 1$ points spaced $\alpha$ apart.

For example, for $p = 1$, the symmetric finite difference stencils are defined as
$$\nabla F_n = \frac{1}{2\alpha}\big(F_{n-1} + F_{n+1}\big) 
\quad\mathrm{and}\quad
\Delta F_n = \frac{1}{\alpha^2}\big(F_{n-1} - 2F_n + F_{n+1} \big),$$
and the $O(\alpha^3)$ term in \eqref{eqn:F_expansion} vanishes. For $p = 2$, we instead define
$$\nabla F_n = \frac{1}{\alpha}\bigg(\frac{1}{12}F_{n-2} - \frac{2}{3}F_{n-1} + \frac{2}{3}F_{n+1} - \frac{1}{12}F_{n+2}\bigg) $$
and
$$\Delta F_n = \frac{1}{\alpha^2}\bigg(-\frac{1}{12}F_{n-2} + \frac{4}{3}F_{n-1} - \frac{5}{2} F_n  + \frac{4}{3}F_{n+1} - \frac{1}{12}F_{n+2} \bigg).$$

\section{Numerical implementation}
\label{app:numerical}

To evolve the amplitude equations \eqref{eqn:dCGLE_intro}, we apply a discrete Fourier transform to $A_n$ and $B_n$, and then introduce an integrating factor to integrate the linear components exactly \cite{milewski1999pseudospectral}. Specifically, we denote the discrete Fourier transform of $A_n$ as $\hat{A}_k = \mathscr{F}_k[A_n]$, and likewise for $B_n$. By applying the Fourier transform to the amplitude equations \eqref{eqn:dCGLE_intro}, we obtain
\begin{subequations}
\label{eqn:dCGLE_app}
\begin{align}
\label{eqn:dCGLE_app1}
\sdone{\hat{A}_k}{T} +  \mathcal{M}_k\hat{A}_k &= \mathscr{F}_k\big[\mu\gamma_{1}A_{n}\nabla B_{n} - \sigma_2 |A_n|^2 A_n\big],\\
\label{eqn:dCGLE_app2} 
\sdone{\hat{B}_k}{T} + \mathcal{N}_k \hat{B}_k &= 
\mathscr{F}_k\big[ 2\mu \text{Re}\left[\gamma_{2}A_{n}\nabla A^{*}_{n}\right] + \mu\gamma_{3} |A_{n}|^2\big],
\end{align}
\end{subequations}
where
$$\mathcal{M}_k =  \mu^2 c_g\nabla_k - \sigma_{1} - \mu^2 D_1\Delta_k \quad \mathrm{and}\quad \mathcal{N}_k = - \mu^2 D_2\Delta_k. $$
We recall that $\nabla_k$ and $\Delta_k$ are the Fourier multipliers of the difference operators $\nabla$ and $\Delta$, respectively. To account for the stiffness manifest in the operators $\mathcal{M}_k$ and $\mathcal{N}_k$, we introduce an integrating factor. When evolving from time $T = T_n$ and $T = T_{n+1}$, we therefore recast \eqref{eqn:dCGLE_app} as
\begin{subequations}
\label{eqn:dCGLE_int}
\begin{align}
\label{eqn:dCGLE_int1}
\sdone{}{T}\big( \hat{A}_k \mye^{\mathcal{M}_k(T - T_n)}\big) &= \mathscr{F}_k\big[\mu\gamma_{1}A_{n}\nabla B_{n} - \sigma_2 |A_n|^2 A_n\big] \mye^{\mathcal{M}_k(T - T_n)},\\
\label{eqn:dCGLE_int2} 
\sdone{}{T} \big(\hat{B}_k \mye^{\mathcal{N}_k(T - T_n)}\big) &= 
\mathscr{F}_k\big[ 2\mu \text{Re}\left[\gamma_{2}A_{n}\nabla A^{*}_{n}\right] + \mu\gamma_{3} |A_{n}|^2\big] \mye^{\mathcal{N}_k(T - T_n)},
\end{align}
\end{subequations}
and then evolve the new variables, $\hat{A}_k \mye^{\mathcal{M}_k(T - T_n)}$ and $\hat{B}_k \mye^{\mathcal{N}_k(T - T_n)}$, using a fourth-order Runge-Kutta method. For the numerical results presented in \S\ref{sec:numerical_solns}, we use a time step of 0.01. MATLAB code implementing this numerical scheme is provided in the supplementary material.
\end{appendices}

\bibliography{gl_bib}

\providecommand{\noopsort}[1]{}\providecommand{\singleletter}[1]{#1}%
\begin{thebibliography}{10}

\bibitem{aranson2002world}
I.~S. Aranson and L.~Kramer.
\newblock The world of the complex {G}inzburg--{L}andau equation.
\newblock {\em Reviews of Modern Physics}, 74(1):99, 2002.

\bibitem{garcia2012complex}
V.~Garc{\'\i}a-Morales and K.~Krischer.
\newblock The complex {G}inzburg--{L}andau equation: an introduction.
\newblock {\em Contemporary Physics}, 53(2):79--95, 2012.

\bibitem{cross1993pattern}
M.~C. Cross and P.~C. Hohenberg.
\newblock Pattern formation outside of equilibrium.
\newblock {\em Reviews of Modern Physics}, 65(3):851, 1993.

\bibitem{newell1993order}
A.~C. Newell, T.~Passot, and J.~Lega.
\newblock Order parameter equations for patterns.
\newblock {\em Annual Review of Fluid Mechanics}, 25(1):399--453, 1993.

\bibitem{coullet1989defect}
P.~Coullet, L.~Gil, and J.~Lega.
\newblock Defect-mediated turbulence.
\newblock {\em Physical Review Letters}, 62(14):1619, 1989.

\bibitem{tanaka2003complex}
D.~Tanaka and Y.~Kuramoto.
\newblock Complex {G}inzburg-{L}andau equation with nonlocal coupling.
\newblock {\em Physical Review E}, 68(2):026219, 2003.

\bibitem{garcia2008nonlocal}
V.~Garc{\'\i}a-Morales and K.~Krischer.
\newblock Nonlocal complex {G}inzburg-{L}andau equation for electrochemical
  systems.
\newblock {\em Physical Review Letters}, 100(5):054101, 2008.

\bibitem{denk2016active}
J.~Denk, L.~Huber, E.~Reithmann, and E.~Frey.
\newblock Active curved polymers form vortex patterns on membranes.
\newblock {\em Physical Review Letters}, 116(17):178301, 2016.

\bibitem{tan2020topological}
T.~H. Tan, J.~Liu, P.~W. Miller, M.~Tekant, J.~Dunkel, and N.~Fakhri.
\newblock Topological turbulence in the membrane of a living cell.
\newblock {\em Nature Physics}, 16(6):657--662, 2020.

\bibitem{newell1969finite}
A.~C. Newell and J.~A. Whitehead.
\newblock Finite bandwidth, finite amplitude convection.
\newblock {\em Journal of Fluid Mechanics}, 38(2):279--303, 1969.

\bibitem{segel1969distant}
L.~A. Segel.
\newblock Distant side-walls cause slow amplitude modulation of cellular
  convection.
\newblock {\em Journal of Fluid Mechanics}, 38(1):203--224, 1969.

\bibitem{cross1980derivation}
M.~C. Cross.
\newblock Derivation of the amplitude equation at the {R}ayleigh--{B}{\'e}nard
  instability.
\newblock {\em Physics of Fluids}, 23(9):1727--1731, 1980.

\bibitem{stoop2015curvature}
N.~Stoop, R.~Lagrange, D.~Terwagne, P.~M. Reis, and J.~Dunkel.
\newblock Curvature-induced symmetry breaking determines elastic surface
  patterns.
\newblock {\em Nature Materials}, 14(3):337--342, 2015.

\bibitem{hakim1992dynamics}
V.~Hakim and W-J. Rappel.
\newblock Dynamics of the globally coupled complex {G}inzburg-{L}andau
  equation.
\newblock {\em Physical Review A}, 46(12):R7347, 1992.

\bibitem{ravoux2000stability}
J.~F. Ravoux, S.~Le~Dizes, and P.~Le~Gal.
\newblock Stability analysis of plane wave solutions of the discrete
  {G}inzburg-{L}andau equation.
\newblock {\em Physical Review E}, 61(1):390, 2000.

\bibitem{maruno2003exact}
K-I. Maruno, A.~Ankiewicz, and N.~Akhmediev.
\newblock Exact localized and periodic solutions of the discrete complex
  {G}inzburg--{L}andau equation.
\newblock {\em Optics Communications}, 221(1-3):199--209, 2003.

\bibitem{thomson2020collective}
S.~J. Thomson, M.~M.~P. Couchman, and J.~W.~M. Bush.
\newblock Collective vibrations of confined levitating droplets.
\newblock {\em Physical Review Fluids}, 5:083601, Aug 2020.

\bibitem{thomson2020hydro}
S.~J. Thomson, M.~Durey, and R.~R. Rosales.
\newblock Collective vibrations of a hydrodynamic active lattice.
\newblock {\em Proceedings of the Royal Society A}, 476(2239), 2020.

\bibitem{sethia2014chimera}
G.~C Sethia and A.~Sen.
\newblock Chimera states: the existence criteria revisited.
\newblock {\em Physical Review Letters}, 112(14):144101, 2014.

\bibitem{thakur2019collective}
B.~Thakur and A.~Sen.
\newblock Collective dynamics of globally delay-coupled complex
  {G}inzburg-{L}andau oscillators.
\newblock {\em Chaos: an Interdisciplinary Journal of Nonlinear Science},
  29(5):053104, 2019.

\bibitem{couder2005walking}
Y.~Couder, S.~Protiere, E.~Fort, and A.~Boudaoud.
\newblock Walking and orbiting droplets.
\newblock {\em Nature}, 437(7056):208--208, 2005.

\bibitem{bush2015pilot}
J.~W.~M. Bush.
\newblock Pilot-wave hydrodynamics.
\newblock {\em Annual Review of Fluid Mechanics}, 47:269--292, 2015.

\bibitem{bush2018introduction}
J.~W.~M. Bush, Y.~Couder, T.~Gilet, P.~A. Milewski, and A.~Nachbin.
\newblock Introduction to focus issue on hydrodynamic quantum analogs.
\newblock {\em Chaos: an Interdisciplinary Journal of Nonlinear Science},
  28(9):096001, 2018.

\bibitem{eddi2011information}
A.~Eddi, E.~Sultan, J.~Moukhtar, E.~Fort, M.~Rossi, and Y.~Couder.
\newblock Information stored in {F}araday waves: the origin of a path memory.
\newblock {\em Journal of Fluid Mechanics}, 674:433--463, 2011.

\bibitem{molavcek2013walking}
J.~Mol{\'a}{\v{c}}ek and J.~W.~M. Bush.
\newblock Drops walking on a vibrating bath: towards a hydrodynamic pilot-wave
  theory.
\newblock {\em Journal of Fluid Mechanics}, 727:612--647, 2013.

\bibitem{bechinger2016active}
C.~Bechinger, R.~Di~Leonardo, H.~L{\"o}wen, C.~Reichhardt, G.~Volpe, and
  G.~Volpe.
\newblock Active particles in complex and crowded environments.
\newblock {\em Reviews of Modern Physics}, 88(4):045006, 2016.

\bibitem{scholz2018inertial}
C.~Scholz, S.~Jahanshahi, A.~Ldov, and H.~L{\"o}wen.
\newblock Inertial delay of self-propelled particles.
\newblock {\em Nature Communications}, 9(1):1--9, 2018.

\bibitem{lim2019cluster}
M.~X. Lim, A.~Souslov, V.~Vitelli, and H.~M. Jaeger.
\newblock Cluster formation by acoustic forces and active fluctuations in
  levitated granular matter.
\newblock {\em Nature Physics}, 15(5):460--464, 2019.

\bibitem{lowen2020inertial}
H.~L{\"o}wen.
\newblock Inertial effects of self-propelled particles: From active {B}rownian
  to active {L}angevin motion.
\newblock {\em The Journal of Chemical Physics}, 152(4):040901, 2020.

\bibitem{oza2013trajectory}
A.~U. Oza, R.~R. Rosales, and J.~W.~M. Bush.
\newblock A trajectory equation for walking droplets: hydrodynamic pilot-wave
  theory.
\newblock {\em Journal of Fluid Mechanics}, 737:552--570, 2013.

\bibitem{strogatz2018nonlinear}
S.~H. Strogatz.
\newblock {\em Nonlinear dynamics and chaos: with applications to physics,
  biology, chemistry, and engineering}.
\newblock CRC press, 2018.

\bibitem{kevorkian2012multiple}
J.~K. Kevorkian and J.~D. Cole.
\newblock {\em Multiple scale and singular perturbation methods}, volume 114.
\newblock Springer Science \& Business Media, 2012.

\bibitem{doering1988low}
C.~R. Doering, J.~D. Gibbon, D.~D. Holm, and B.~Nicolaenko.
\newblock Low-dimensional behaviour in the complex {G}inzburg-{L}andau
  equation.
\newblock {\em Nonlinearity}, 1(2):279, 1988.

\bibitem{bartuccelli1990possibility}
M.~Bartuccelli, P.~Constantin, C.~R. Doering, J.~D. Gibbon, and
  M.~Gisself{\"a}lt.
\newblock On the possibility of soft and hard turbulence in the complex
  {G}inzburg-{L}andau equation.
\newblock {\em Physica D: Nonlinear Phenomena}, 44(3):421--444, 1990.

\bibitem{coullet1985propagative}
P.~Coullet and S.~Fauve.
\newblock Propagative phase dynamics for systems with {G}alilean invariance.
\newblock {\em Physical Review Letters}, 55(26):2857, 1985.

\bibitem{matthews2000pattern}
P.~C. Matthews and S.~M. Cox.
\newblock Pattern formation with a conservation law.
\newblock {\em Nonlinearity}, 13(4):1293, 2000.

\bibitem{komarova2000nonlinear}
N.~L. Komarova and A.~C. Newell.
\newblock Nonlinear dynamics of sand banks and sand waves.
\newblock {\em Journal of Fluid Mechanics}, 415:285--321, 2000.

\bibitem{cox2001new}
S.~M. Cox and P.~C. Matthews.
\newblock New instabilities in two-dimensional rotating convection and
  magnetoconvection.
\newblock {\em Physica D: Nonlinear Phenomena}, 149(3):210--229, 2001.

\bibitem{benjamin1967disintegration}
T.~B. Benjamin and J.~E. Feir.
\newblock The disintegration of wave trains on deep water.
\newblock {\em Journal of Fluid Mechanics}, 27(3):417--430, 1967.

\bibitem{stuart1978eckhaus}
J.~T. Stuart and R.~C. DiPrima.
\newblock The {E}ckhaus and {B}enjamin-{F}eir resonance mechanisms.
\newblock {\em Proceedings of the Royal Society of London. A: Mathematical and
  Physical Sciences}, 362(1708):27--41, 1978.

\bibitem{milewski1999pseudospectral}
P.~A. Milewski and E.~G. Tabak.
\newblock A pseudospectral procedure for the solution of nonlinear wave
  equations with examples from free-surface flows.
\newblock {\em SIAM Journal on Scientific Computing}, 21(3):1102--1114, 1999.

\bibitem{durey2020faraday}
M.~Durey, P.~A. Milewski, and Z.~Wang.
\newblock Faraday pilot-wave dynamics in a circular corral.
\newblock {\em Journal of Fluid Mechanics}, 891:A3, 2020.

\bibitem{abrams2004chimera}
D.~M. Abrams and S.~H. Strogatz.
\newblock Chimera states for coupled oscillators.
\newblock {\em Physical Review Letters}, 93(17):174102, 2004.

\bibitem{sethia2008clustered}
G.~C. Sethia, A.~Sen, and F.~M. Atay.
\newblock Clustered chimera states in delay-coupled oscillator systems.
\newblock {\em Physical Review Letters}, 100(14):144102, 2008.

\bibitem{nkomo2013chimera}
S.~Nkomo, M.~R. Tinsley, and K.~Showalter.
\newblock Chimera states in populations of nonlocally coupled chemical
  oscillators.
\newblock {\em Physical Review Letters}, 110(24):244102, 2013.

\bibitem{martens2013chimera}
E.~A. Martens, S.~Thutupalli, A.~Fourri{\`e}re, and O.~Hallatschek.
\newblock Chimera states in mechanical oscillator networks.
\newblock {\em Proceedings of the National Academy of Sciences},
  110(26):10563--10567, 2013.

\bibitem{wojewoda2016smallest}
J.~Wojewoda, K.~Czolczynski, Y.~Maistrenko, and T.~Kapitaniak.
\newblock The smallest chimera state for coupled pendula.
\newblock {\em Scientific Reports}, 6(1):1--5, 2016.

\bibitem{totz2018spiral}
J.~F. Totz, J.~Rode, M.~R. Tinsley, K.~Showalter, and H.~Engel.
\newblock Spiral wave chimera states in large populations of coupled chemical
  oscillators.
\newblock {\em Nature Physics}, 14(3):282--285, 2018.

\end{thebibliography}
\bibliographystyle{unsrt}

\end{document}